\documentclass[a4paper,aps,prd,twocolumn,showpacs,eqsecnum,amsmath,amssymb,nofootinbib,superscriptaddress]{revtex4-1}
\usepackage{dcolumn}
\usepackage{bm}
\usepackage{graphics}
\usepackage{graphicx}
\usepackage{epsfig}
\usepackage{booktabs,multirow}
\usepackage{hhline}
\usepackage{slashed}
\usepackage{rccol}
\usepackage{mathrsfs}
\usepackage{xcolor,colortbl}

\usepackage{xcolor}\colorlet{linkequation}{blue}
\usepackage[colorlinks=true, 
            linkcolor=red,   
            filecolor=red,   
            citecolor=green, 
            urlcolor=blue,
            hyperfootnotes=true]{hyperref}

\newcommand*{\SavedEqref}{}
\let\SavedEqref\eqref
\renewcommand*{\eqref}[1]{%
  \begingroup
    \hypersetup{
     linkcolor=linkequation,
      linkbordercolor=linkequation,
    }%
    \SavedEqref{#1}%
  \endgroup
}

\begin{document}

\def\beqa{\begin{eqnarray}}
\def\eeqa{\end{eqnarray}}
\newcommand{\be}{\ensuremath{\beta}}
\newcommand{\al}{\ensuremath{\alpha}}
\newcommand{\sa}{\ensuremath{\sin\alpha}}
\newcommand{\ca}{\ensuremath{\cos\alpha}}
\newcommand{\ta}{\ensuremath{\tan\alpha}}
\newcommand{\sbt}{\ensuremath{\sin\beta}}
\newcommand{\cbt}{\ensuremath{\cos\beta}}
\newcommand{\cba}{\ensuremath{c_{\beta-\alpha}}}
\newcommand{\ma}{\ensuremath{m_{A}}}
\newcommand{\mh}{\ensuremath{m_{h^0}}}
\newcommand{\mH}{\ensuremath{m_{H^0}}}
\newcommand{\mev}{\mbox{~MeV}}
\newcommand{\gev}{\mbox{~GeV}}
\newcommand{\tev}{\mbox{~TeV}}

\newcommand{\ben}{\begin{enumerate}}
\newcommand{\een}{\end{enumerate}}
\newcommand{\bc}{\begin{center}}
\newcommand{\ec}{\end{center}}
\newcommand{\mb}{\mbox{\ }}
\newcommand{\vs}{\vspace}
\newcommand{\ra}{\rightarrow}
\newcommand{\la}{\leftarrow}
\newcommand{\ul}{\underline}
\newcommand{\ds}{\displaystyle}
\definecolor{LightCyan}{rgb}{0.0, 1, 0.94}

\title{Precise predictions for charged Higgs boson pair production in photon-photon collisions}

\author{Mehmet~Demirci}
\email{mehmetdemirci@ktu.edu.tr}
\affiliation{Department of Physics, Karadeniz Technical University, 61080 Trabzon, Turkey}%
\date{\today}
\begin{abstract} The charged Higgs pair production via photon-photon collisions is investigated in the framework of two Higgs doublet model (2HDM), taking into account a full set of one-loop-level scattering amplitudes, i.e., including electroweak (EW) corrections together with soft and hard QED radiation. The numerical evaluation is carried out for three different scenarios, so-called non-alignment, low-$m_H$ and short-cascade, defined in the presence of the up-to-date experimental constraints and consistent with theoretical constraints as well. The total cross sections of $\gamma\gamma \rightarrow H^{-}H^{+}$ are scanned over the plane ($m_{\phi^0},\sqrt{s}$), where $\phi^0$ is $h^0$ for low-$m_{H^0}$ scenario and $H^0$ for other two scenarios.  The regions of the parameter space in which the production rates are sufficiently large are highlighted for each scenario. The production rates in different polarization collision modes of initial beams are also discussed. It can be enhanced up to two-times by oppositely polarized photons at high energies and right-handed polarized photons at low energies. Furthermore, decay channels of the charged Higgs boson are examined for each scenario. It is observed that the one-loop EW corrections,  i.e., the virtual plus real radiation corrections, reduce the tree-level cross sections and the relative correction is typically in the range of $-10\%$ to $-30\%$, depending on model parameters.
\end{abstract}


\maketitle

\section{Introduction}
The Standard Model (SM) is the most successful model of particle physics so far, because its theoretical predictions are magnificently compatible with experiments. The electroweak symmetry breaking in the SM is successfully accomplished through the Higgs mechanism, which yields their masses to the fundamental particles and suggests the existence of a Higgs boson. A huge step was taken for high energy physics with a 125 GeV of Higgs boson observed by both CMS and ATLAS experiments at the LHC~\cite{ATLAS,CMS,ATLASCMS}. Although the properties of the Higgs boson discovered are suitable with the SM predictions, it is possible that it constitutes but one state of a richer Higgs sector. In association with that, the models with extended Higgs sector are among the best motivated beyond SM models (BSM), as they may provide solutions to many deficiencies in the SM, such as the gauge hierarchy problem, the origin of dark matter, the generation of a baryon asymmetry and the strong CP problem.

One of the simplest such extensions is the two-Higgs doublet model (2HDM)~\cite{THDM00,THDM12} that only adds one more Higgs doublet to the SM under the same (SM) gauge symmetry. Its scalar sector consists of five scalars, one CP-odd Higgs (or pseudoscalar) $A^0$, two CP-even Higgs bosons ($h^0$ and $H^0$) and two charged Higgs $H^\pm$. Versions of the 2HDM are often used in several of well-founded new physics BSM scenarios, both with and without Supersymmetry~\cite{Haber1985}, where the additional Higgs doublet is either an essential by-product or a necessary component in indicating issues.

Since it has an ultimate point as a high-precision experimental machine and hence its complementary with LHC, it is useful to investigate the phenomenology of the Higgs sector in detail in the context of an electron-positron collider. In this respect, the clean environment in these colliders would ensure that the Higgs sector is precisely identified. International Linear Collider (ILC)~\cite{ILC1,ILC2} is an efficient machine for precise experiments which is aimed to yield equipment for electron-positron collisions along with other possibilities such as electron-electron, electron-photon and photon-photon collisions. The Compact Linear Collider (CLIC)~\cite{CLIC1} is a TeV–scale high-luminosity linear collider operating with the center-of-mass energy energy up to $3\tev$. The primary task of these linear colliders is to expand and complement the results obtained in the hadron colliders, and to discover new physics BSM. Besides, the photon-photon-collision is of the opinion as other collision mode which can yield an integrated luminosity of the order of one hundred fb$^{-1}$ per year. This collision can produce unique new physics compared to other types of collisions. Upgrading the machine is expected to reach the high energies $\sqrt{s} = 1\tev$ with up to three hundred fb$^{-1}$ per year of total integrated luminosity~\cite{ILC3}. In addition to the possible discovery of the Higgs boson relatives through examining the observed-Higgs boson properties, the ILC presents great opportunities to explore new lighter Higgs bosons –or, more commonly, any weakly interacting light (pseudo)scalar boson– via the direct production~\cite{Fujii2017}. Discovering the charged Higgs boson would become a clear proof of physics BSM and an eminent signal for extended Higgs sector. An extensive review on charged Higgs phenomenology
is available in Ref.~\cite{Akeroyd2017}. The future $e^-e^+$ and $\gamma\gamma$ colliders with high energy and high luminosity have significant potential in the discovery of the charged Higgs boson and in the study of its properties. Furthermore, high energy $\gamma\gamma$-collisions, where photons directly coupled to charged particles, can give a better understanding of the SM and its extensions for several aspects. Correspondingly, exact predictions are needed for the physical observables related to charged Higgs bosons.

The main channels of the pair production for charged Higgs bosons in the linear colliders are $e^- e^+ \rightarrow H^{-}H^{+}$ and $\gamma\gamma \rightarrow H^{-}H^{+}$. Since the cross-section of $e^- e^+ \rightarrow H^{-}H^{+}$ is suppressed by s-channel contributions, the production rate of $\gamma\gamma \rightarrow H^{-}H^{+}$ is larger than that of $e^- e^+$-collision mode. The scattering process $e^- e^+ \rightarrow H^{-}H^{+}$ has been widely studied by including the one-loop corrections in the framework of both 2HDM and MSSM~\cite{Komamiya88, Arhrib95,Arhrib99, Moretti03,Hashemi14,Heinemeyer16}. On the other hand, the scattering process $\gamma\gamma \rightarrow H^{-}H^{+}$ has been studied at the one-loop level by including the full squark corrections~\cite{Zhu98} and the real radiative corrections~\cite{Lei05,Sonmez20} in the framework of SUSY but only yukawa corrections in the 2HDM~\cite{Liang96}. Full one-loop level corrections for this production mode should be considered in the framework of the 2HDM. In order to benefit from the high precision measurements, we also need a high precision predictions from the theory, which means that there is a need to go beyond the leading order calculations for most processes, hence the full one-loop contributions become significant for physics analyses at the future colliders, e.g. ILC or CLIC. The scattering process $\gamma\gamma \rightarrow H^{-}H^{+}$ have rich physics results and needs a detailed study in the light of current constraints.

In the present work, the production of the charged Higgs bosons pairs through photon-photon collision is investigated in the framework of 2HDM for the first time, including the full one-loop contributions, i.e. electroweak (EW) corrections, as well as hard and soft QED radiation. For numerical evaluation, six different benchmark points, which have a
\textit{CP}-even scalar with mass of 125 GeV and couplings compatible with those of the observed Higgs boson, are chosen. They are constructed from the scenarios so-called as ``non-alignment'', ``low-$m_H$" and ``short-cascade"~\cite{BPscenarios}. These scenarios are defined in the presence of the up-to-date experimental constraints and consistent with theoretical constraints as well. The total cross sections of $\gamma\gamma \rightarrow H^{-}H^{+}$ are scanned over the plane ($m_{\phi^0},\sqrt{s}$), where $\phi^0$ is $h^0$ for low-$m_{H^0}$ scenario and $H^0$ for other two scenarios. The parameter space regions in which the improvement of the production rate is significant enough to be accessible at the future colliders are highlighted for each scenario. The production rates in different polarization collision modes of initial beams are also discussed.

The rest of this work is organized as follows. In Section \ref{sec:THDM}, a brief review of the 2HDM is presented. In Sec.~\ref{sec:parameter}, we review the theoretical and experimental constraints imposed on the 2HDM. Then, the benchmark scenarios used in the calculation are given. In Sec.~\ref{sec:cros}, the Feynman diagrams, the corresponding amplitudes, and some useful analytical expressions are given for the considered scattering process. The general forms of the virtual and the real photon radiation corrections are also discussed. In Sec.~\ref{sec:results},  we present the numerical results related to the scattering process and decay channels, and discuss in detail the corresponding model parameter dependencies of the cross sections.  Finally, the concluding remarks are presented in Sec.~\ref{sec:conc}.

\section{Review of the two Higgs doublet model}\label{sec:THDM}
In this section, a brief overview of the 2HDM with CP-conserving, concerning only relevant details for this study, is introduced. The 2HDM is constructed by augmenting the complex scalar doublet, $\Phi_1$, of the SM by another doublet, $\Phi_2$, which changes the dynamics of EW symmetry-breaking. The most general scalar potential being invariant under the ${\rm SU(2)_L}\otimes  {\rm  U(1)_Y}$ gauge group is defined by
\begin{equation} \label{eq:potential}
\begin{split}
V_{\text{2HDM}}  = & m_1^2 |\Phi_1|^2 +m_2^2 |\Phi_2|^2
-\bigg[m_{12}^2 (\Phi_1^\dag \Phi_2) \\ & +{\rm h.c.}\bigg]
+\frac{\lambda_1}{2}(\Phi_1^\dag \Phi_1)^2+\frac{\lambda_2}{2}(\Phi_2^\dag \Phi_2)^2\\
& +\lambda_3(\Phi_1^\dag \Phi_1)(\Phi_2^\dag \Phi_2)
 +\lambda_4(\Phi_1^\dag \Phi_2)(\Phi_2^\dag \Phi_1) \\
&+ \bigg[\frac{\lambda_5}{2}(\Phi_1^\dag \Phi_2)^2
+\lambda_6(\Phi_1^\dag \Phi_1)(\Phi_1^\dag \Phi_2)\\
&+\lambda_7(\Phi_1^\dag \Phi_2)(\Phi_2^\dag \Phi_2)+{\rm h.c.}\bigg]
\end{split}
\end{equation}
where $\lambda_i$ (i=1,...,7) are quartic coupling parameters\footnote{Note that the parameters in~\eqref{eq:potential} have to be real for the CP-conservation.} and  $\Phi_{1,2}$ are two complex scalar Higgs doublets. To comply with some low energy observables, the discrete $Z_2$ symmetry is put forward by the Paschos-Glashow-Weinberg theorem~\cite{Glashow,Paschos}. In particular, this symmetry is applied to prevent flavor changing neutral currents at tree-level. The $Z_2$ symmetry requires to $\lambda_{6,7}$ and $m_{12}^2$ be zero. However, allowing $m_{12}^2$ be non-zero, the $Z_2$ symmetry are softly broken. The charges assigned under the $Z_2$ symmetry allow that each type of fermion interaction with only one Higgs doublet, i.e., $\Phi_{1}$  or $\Phi_{2}$. According to the $Z_2$ assignment, there appear four 2HDM-types, which are mostly categorized as type-I, II, III (or ``lepton-specific'') and IV (or ``flipped'') of 2HDM~\cite{THDM00,THDM12}. Table~\ref{tab:coupling} shows of how fermions couple to each Higgs doublet ($\Phi_{1,2}$) in the allowed types where flavor conservation is naturally obeyed. This work is concentrated on the Type-I and Type-II of 2HDM. In Type-I, only the doublet $\Phi_2$ interacts with both leptons and quarks like in the SM. In Type-II, the doublet $\Phi_1$ couples to leptons and $d$-type quarks, while $\Phi_2$ couples to $u$-type quarks.
\begin{table}[h]
\caption{Higgs doublets $\Phi_{1,2}$ couplings to $u$-type quarks, $d$-type quarks and charged leptons in the four different types of 2HDMs allowed by the $Z_2$ symmetry.
\label{tab:coupling}}
\begin{ruledtabular}
  \begin{tabular}{rlll}
type& $u_i$ & $d_i$ & $\ell_i$ \\
\hline
  I & $\Phi_2$ & $\Phi_2$& $\Phi_2$ \\
 II & $\Phi_2$ & $\Phi_1$& $\Phi_1$ \\
III & $\Phi_2$ & $\Phi_2$& $\Phi_1$  \\
 IV & $\Phi_2$ & $\Phi_1$& $\Phi_2$ \\
\end{tabular}
\end{ruledtabular}
\end{table}

After the spontaneous symmetry breaking of the ${\rm SU(2)}_L\otimes {\rm  U(1)}_Y$ gauge symmetry associated with the electroweak force, the neutral components of scalar doublet acquire vacuum expectation values $v_j$ such that
\begin{eqnarray}
\Phi_j  =  \left(\begin{array}{c}
  \phi_j^+ \\  \frac{1}{\sqrt{2}}(v_j+\rho_j+i \eta_j) \end{array}\right), (j=1,2),
\end{eqnarray}
where $\rho_j$ and $\eta_j$ are real scalar fields. The combination $v^2=v_1^2+v_2^2 \simeq (246 \gev)^2$ is set by
its relation to the mass of $W$ and the Fermi constant as follows: $v^2=1/(\sqrt{2} G_F)=4m_W^2/g^2$. These Higgs doublets include initially eight degrees of freedom. The three of them, $G^{\pm}$, $G^0$ bosons, are eaten by the longitudinal components of the EW vector bosons $Z$ and $W^\pm$. The remaining ones are five physical Higgs bosons: two CP-even $h^0$ and $H^0$, a CP-odd Higgs $A^0$,  and two charged scalars $H^\pm$. They are related to the weak eigenstates via
\begin{equation}\label{eq:higgmasseigen}
\begin{split}
&\left(\begin{array}{c}
\phi_1^\pm\\
\phi_2^\pm
\end{array}\right)=R_\beta
\left(\begin{array}{c}
G^\pm\\
H^\pm
\end{array}\right),
\left(\begin{array}{c}
\rho_1\\
\rho_2
\end{array}\right)=R_\alpha
\left(\begin{array}{c}
H^0\\
h^0
\end{array}\right) ,
\\
&\left(\begin{array}{c}
\eta_1\\
\eta_2
\end{array}\right)
=R_\beta\left(\begin{array}{c}
G^0\\
A^0
\end{array}\right),
\end{split}
\end{equation}
with the generic rotation matrix\footnote{Here and in the following, the short-hand notation $c_x \equiv cos(x)$, $s_x \equiv sin(x)$ and $t_x \equiv tan(x)$ will be used.}
$$R_\theta =
\left(
\begin{array}{cc}
c_{\theta} & -s_{\theta}\\
s_{\theta} & c_{\theta}
\end{array}\right). $$

For any given value of $t_\beta$, the parameters $m_1^2$ and $m_2^2$ are determined by the minimization conditions of potential. The mass parameters $m^2_{1,2}$ and quartic couplings $\lambda_1$--$\lambda_5$ can be defined in terms of the physical masses $m_h$, $m_H$, $m_A$, $m_{H^\pm}$, along with $t_\beta=v_2/v_1$ (the ratio of vacuum expectation values), and the mixing term of neutral sector $s_{\beta-\alpha}$. The soft $Z_2$ symmetry breaking parameter $m^2_{12}$ is defined by
\begin{equation}\label{eq:m122}
m^2_{12}=\frac{1}{2} \lambda_5 v^2 s_\beta c_\beta = \frac{\lambda_5}{2\sqrt{2} G_F} \frac{t_\beta}{1+t^2_\beta},
\end{equation}
where the last equality is only for the tree level. Fixing $\lambda_{6}$ and $\lambda_{7}$ to zero to respect the $Z_2$ symmetry, $m_{12}^2$, $\tan\beta$, mixing angle $\alpha$ and four Higgs boson masses are enough to specify the model completely in the physical basis \footnote{There are also other bases constructed with different parametrizations of the 2HDM potential, which are so-called the Higgs basis and the hybrid basis.}. Consequently, there are seven independent free parameters encountered in the Higgs sector of the 2HDM.

The phenomenology of the 2HDM depends significantly on the size of the mixing angle $\alpha$ together with angle $\beta$. There appears an alignment limit~\cite{Carena}, where the CP-even Higgs boson $h^0$ looks SM-like Higgs boson if $s_{\beta-\alpha}\to 1$ or $c_{\beta-\alpha}\to 0$. The alignment limit is the most favoured condition by the experimentalists. In this limit, the couplings of the \textit{CP}-even Higgs boson $h^0$ in 2HDM are similar to Higgs boson of SM and can be identified as the discovered 125 GeV Higgs boson. Furthermore, the \textit{CP}-even Higgs boson $H^0$ acts as gauge-phobic such that its coupling to the vector bosons $Z/W^\pm$ is greatly suppressed. However, in the limit $c_{\beta-\alpha}\to 1$, $H^0$ looks SM-like Higgs boson.

Furthermore, a decoupling limit appears when $c_{\beta-\alpha}=0$ and $m_{H^0,A^0,H^\pm}\gg m_Z$ \cite{Gunion}. At this limit, the Higgs boson $h^0$ coupling to SM particles completely appear like the couplings of the SM-Higgs boson  that contain the coupling $h^0h^0h^0$.

\section{Parameter Setting on 2HDM}\label{sec:parameter}
The parameter space of 2HDM is subjected to both the bounds coming from experimental searches and theoretical constraints that arise from the model itself. These have to be imposed to the free parameters of the model.

\subsection{Theoretical and experimental constraints}
The parameter space of the scalar 2HDM potential is reduced by the theoretical constraints: potential stability, perturbativity and unitarity. The ${V}_{\rm 2HDM}$ is bounded from below respect to the vacuum stability of the 2HDM. Namely, ${V}_{\rm 2HDM}\geq 0$ must be satisfied for all directions of $\Phi_1$ and $\Phi_2$. Consequently, the following conditions are placed on the parameters $\lambda_i$ \cite{BFB1,BFB2,Gunion}:
\begin{equation}
\begin{split}
&\lambda_1>0, ~~\lambda_2>0,\\
&\lambda_3 + \sqrt{\lambda_1 \lambda_2}+ \text{Min}(0, \lambda_4 - |\lambda_5|) > 0.
\end{split}
\end{equation}

Another set of constraints enforce that the perturbative unitarity  (for details, see Refs.~\cite{unitarity1,unitarity2}) need to be fulfilled for scattering of longitudinally polarized gauge and Higgs bosons. Besides, the scalar potential needs to be perturbative by demanding that all quartic coefficients satisfy $|\lambda_{1,2,3,4,5}| \leq 8 \pi$.
Furthermore, note that the global fit to EW measurements dictates $\Delta \rho$ to be $\mathcal{O}(10^{-3})$~\cite{PDG,EWPO}. This forbids the occurrence of large mass splitting between Higgs bosons of 2HDM, and imposes that $m_{H^\pm}\sim m_A$ or $m_H$ or $m_h$.

Besides the above theoretical constraints, 2HDMs have been researched in the past and still ongoing experiments, such as direct observations at the LHC or indirect $B$~physics observables. Consequently, many results have been accumulated since then, and the parameter space of the 2HDM is restricted by all conducted results. In the Type-I of 2HDM, the following pseudoscalar Higgs mass regions $310<m_A<410$ GeV for $m_H = 150$ GeV, $335<m_A<400$ GeV for $m_H = 200$ GeV, $350<m_A<400$ GeV for $m_H = 250$ GeV with $t_\beta = 10$ have been excluded by the LHC experiment~\cite{ATLAS2}. Furthermore, the CP-odd Higgs mass is bounded as $m_A > 350$ for $t_\beta < 5$~\cite{ATLAS3} and the mass range $170<m_H<360$ GeV with $t_\beta < 1.5$ is excluded for the Type-I~\cite{ATLAS4}.

The charged Higgs boson mass is subject to a number of limits from direct experimental researches at the LHC (and previous colliders) as well as from various $B$-physics observables. In type-II and IV of 2HDM, the $BR(b\to s\gamma)$ measurement places bounds on the charged Higgs mass as $m_{H^\pm}>580$~GeV for $t_\beta \geq 1$~\cite{Misiak2017,Misiak2015}. On the contrary, in the type-I and III of 2HDM, this bound is much lower~\cite{Enomoto:2015wbn}. Considering $t_\beta\geq 2$, there is a possibility for the charged Higgs bosons in type-I and III of 2HDM to be as light as 100 GeV \cite{Enomoto:2015wbn,Arhrib:2016wpw} while at the same time compatible with LEP and LHC limits as well as with constraints of all $B$ physics \cite{Aad:2014kga,Khachatryan:2015qxa,Khachatryan:2015uua,
Aad:2013hla, Abbiendi:2013hk,Akeroyd:2016ymd}. Additionally, according to LHC, Tevatron and LEP results~\cite{Moretti}, there is no exclusion about $s_{\beta-\alpha}= 1$ for $m_{A,H,H^\pm} = 500$ GeV in the Type-I 2HDM.

\subsection{Benchmark points scenarios}
Let us now provide details of the benchmark scenarios
used in this study. The three different benchmark scenarios, which are named non-alignment, low-$m_H$ and short-cascade, proposed in Ref.~\cite{BPscenarios}, are used to investigate the phenomenology of the charged Higgs boson. All of these scenarios contain a $CP$-even scalar with 125 GeV mass and its couplings consistent with the successfully observed Higgs boson. In addition, a significant portion of their parameter space is allowed by the constraints from the extra Higgs bosons searches. The four benchmark points (BPs) are selected from these scenarios as shown in Table~\ref{BP}, and they are agree with experimental and theoretical constraints. For each BP, the quantities of potential stability, perturbativity and unitarity have been verified by using {\tt 2HDMC 1.7.0}~\cite{2HDMC,2HDMC2}.
\begin{table*}[t]
\caption{A list of the input parameters for benchmark scenarios which can be used to realize the 2HDM  in hybrid base.}\label{BP}
\begin{ruledtabular}
\begin{tabular}{lcccccccccc}
Scenario&BPs& $\mh$ (GeV) & $\mH$ (GeV) & $\cba$ & $Z_4$ & $Z_5$ & $Z_7$ & $\tan\beta$ & Type &mass hierarchy\\
\hline
\multirow{1}{*}{\bf Non-alignment}&BP1 & 125 & $150\ldots 600$  & $0.1$ & $-2$ & $-2$ & $0$ &  $1\ldots 50$ & I& $m_{H^0}<m_{H^\pm}=m_{A^0}$\\
\hline
\multirow{2}{*}{\bf Short cascade}
&BP2 &  \multirow{2}{*}{$125$} &  \multirow{2}{*}{$250\ldots 500$} & $0$ & $-1$ & $1$ & $-1$ &\multirow{2}{*}{ $2$} & \multirow{2}{*}{I}& $ m_{A^0}  <m_{H^\pm} = m_{H^0}$  \\
&BP3 &                          &    & $0$ & $2$ & $0$ & $-1$ &   & & $m_{H^\pm} < m_{A^0} = m_{H^0}$\\
\hline
\multirow{1}{*}{\bf Low-$m_H$}
&BP4 & $65 \ldots 120$ &  125 & $1.0$ & $-5$ & $-5$ & $0$ &  1.5 & II&$m_{h^0}<m_{H^\pm}=m_{A^0}$ \\
\end{tabular}
\end{ruledtabular}
\end{table*}

These benchmark scenarios are constructed on the hybrid basis\footnote{For a more detailed explanation of
their phenomenological relevance, see Ref.~\cite{BPscenarios}.} where the input parameter is designated as $\{m_h,m_H, c_{\beta-\alpha}, \tan\beta, Z_4, Z_5, Z_7\}$, for the case of softly broken $Z_2$-invariant 2HDM. Here, the parameters $Z_4, Z_5, Z_7$ are quartic couplings in the Higgs basis of $\mathcal{O}(1)$. The masses of the pseudoscalar Higgs and charged Higgs boson in this basis are given by
\begin{equation}
\begin{split}\label{eq:mcHmA}
&m_{A^0}^2=m_{H^0}^2 s_{\beta-\alpha}^2+m_{h^0}^2 c_{\beta-\alpha}^2-Z_5 v^2,\\
&m_{H^\pm}^2=m_{A^0}^2-\frac{1}{2} (Z_4-Z_5)v^2.
\end{split}
\end{equation}

\begin{itemize}
\item In the \textit{non-alignment scenario}, the discovered Higgs boson is interpreted as the lightest $CP$-even scalar $h^0$, with SM-like properties. In the so-called alignment limit, the $c_{hVV}$ coupling  become the same values as in the SM. In this case, the heavier CP-even Higgs boson $H^0$ could not decay into the gauge bosons $W^-W^+$ and $Z^0Z^0$. However, this scenario is defined with a non-alignment ($c_{\beta-\alpha}\neq 0$) as permitted by the current constraints. This leads to some interesting phenomenology of the $H^0$. The other two Higgs bosons $H^\pm$ and $A^0$ are decoupled as $m_{h^0}=125\gev<m_{H^0}<m_{H^\pm}=m_{A^0}$. In this scenario, to obtain values of $m_{H^\pm}$ which satisfy the $b\to s\gamma$ constraint, quartic coupling parameters are set as $Z_4=Z_5=-2$. Consequently, $\tan\beta$ and $m_{H^0}$ remain as free parameters. One benchmark point, BP1, is selected from this scenario by fixing $c_{\beta-\alpha}=0.1$ with type I.

  \item The \textit{short-cascade scenario} is established with a SM-like $h^0$ by taking exact alignment, $c_{\beta-\alpha}=0$. The mass hierarchy could be arranged such that it is possible for either one or both of the decay channels $ H^0 \to W^\pm H^\mp$ or $H^0 \to Z^0 A^0$ to be open, and resulting Higgs-to-Higgs decays in a “small cascade”. In this scenario, two of the Higgs masses $m_{A^0}, m_{H^0}, m_{H^\pm}$ are choosen to be equal for simplicity. These mass degeneracies could be arranged by choosing $Z_4$ and $Z_5$ properly. In this study, two different mass hierarchy are considered for Type-I 2HDM. They are as follows: $ m_{A^0}  <m_{H^\pm} = m_{H^0}$ for $Z_4 = -Z_5 = -1$, and $m_{H^\pm} < m_{A^0} = m_{H^0}$ for $Z_4 = 2$ and $Z_5 = 0$, while keeping the remaining parameters fixed such as $\tan\beta = 2$, $Z_7=-1$. Two benchmark points, BP2 and BP3, are taken from this scenario by fixing $c_{\beta-\alpha}=0$ with type-I.

  \item Another scenario is the \textit{low-$m_H$ scenario} where both $CP$-even Higgs bosons ($h^0$, $H^0$) are light, however; the heavier one ($H^0$) is assumed as SM-like Higgs boson so that $m_{H^0} = 125\gev$. The coupling of the heavier CP-even Higgs to gauge bosons is proportional to $c_{\beta-\alpha}$. Because $m_h < m_H$, the couplings of lighter CP-even scalars to vector bosons must have been heavily suppressed to agree with direct search bounds which forces $s_{\beta-\alpha} \to 0$. Similarly to non-alignment scenario, to decouple the Higgs bosons $A^0$ and $H^\pm$, the quartic parameters $Z_4$ and $Z_5$ are taken as $Z_4=Z_5=-5$. The corresponding mass hierarchy is $m_{h^0}<m_{H^0}=125\gev<m_{H^\pm}=m_{A^0}$. The parameter space for $90 < m_h < 120\gev$ is restricted by the searches $h \rightarrow bb, \tau\tau$ at the LHC, leads to an upper bound on $\tan\beta$~\cite{BPscenarios}. Therefore, it is taken as $\tan\beta = 1.5$. The BP4 is selected from this scenario by fixing $c_{\beta-\alpha}=1.0$ with type II.
\end{itemize}

\section{The cross section of the charged Higgs boson pair production in photon-photon collisions: Theoretical Setup}\label{sec:cros}
The future $e^{-}e^{+}$ colliders could also supply another possibility to measure the production of $H^{-}H^{+}$ in photon-photon collision mode. This can be carried through the Compton backscattered photons, produced in the intense laser photon scatterings on the initial electron-positron beams. Similar with the study on the scattering process $e^{-}e^{+} \rightarrow H^{-}H^{+}$, the evaluation of the corresponding pair production in photon-photon collisions is also crucial for the precise experimental measurements of production of $H^{-}H^{+}$ at an $e^{-}e^{+}$ linear collider.

Here, some details about the calculation of the tree-level and one-loop corrections of cross section are respectively provided. The relevant production channel is expressed by
\begin{equation} \label{eq:gammagammaHH}
\gamma(p_1,\lambda_1)\gamma(p_2,\lambda_2)\rightarrow H^{-}(k_1) H^{+}(k_2),
\end{equation}
where the quantities $p_1,p_2,k_1$ and $k_2$ in parentheses label the corresponding particle 4-momenta. $\lambda_1$ and $\lambda_2$ denote the helicities of initial photons and can take the value of either $+1$ or $-1$, corresponding to a right-handedly (R) and left-handedly (L) polarized photon beam, respectively. All these momenta obey the on-shell equations $p_1^2=p_2^2=0$ and $k_1^2=k_2^2=m_{H^\pm}^2$. Also note for further use the Mandelstam variables:
\begin{equation}
\begin{split}
\hat{s}=(k_1+k_2)^2=(p_1+p_2)^2, \\
\hat{t}=(k_1-p_1)^2=(k_2-p_2)^2, \\
\hat{u}=(k_2-p_1)^2=(k_1-p_2)^2.
\end{split} \label{eq:manvar}
\end{equation}
The polarization vectors of the photons are introduced by
\begin{equation}
\begin{split}
\varepsilon^1_{\mu}(p_1,\lambda_1=\pm1)&=-\frac{1}{\sqrt{2}}(0,1,\mp i,0),\\
\varepsilon^2_{\mu}(p_2,\lambda_2=\pm1)&=\frac{1}{\sqrt{2}}(0,1,\pm i,0),
\end{split}
\end{equation}
which ensure $\varepsilon^{i}\cdot p_j=0$ for $i,j=1,2$.

The analytical and numerical evaluation have been obtained by adopting Mathematica packages\footnote{
We have already done several recent works~\cite{Demirci16,Demirci19a,Demirci19b} by using the same tools and achieved significant results.} as follows:
The Feynman diagrams and the relevant amplitudes have been created with the help of \textsc{FeynArts}~\cite{Feynarts}. Then, the squaring the corresponding amplitudes, the simplification
of the fermion chains, and the numerical evaluation have been carried out by \textsc{FormCalc} \cite{Hahn}.  The scalar loop integrals have been evaluated by \textsc{LoopTools}~\cite{loop}.
The phase-space integration is computed by the Monte-Carlo integration algorithm Vegas as implemented in the \textsc{CUBA} library~\cite{CUBA}.

\subsection{Leading Order Calculation}
\begin{figure}[hbt]
    \begin{center}
\includegraphics[width=0.95\linewidth]{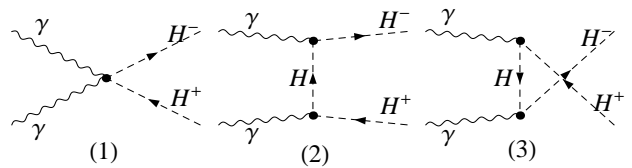}
     \end{center}
          \vspace{-4mm}
\caption{The tree-level Feynman diagrams for the process $\gamma\gamma \rightarrow H^{-}H^{+}$.}\label{fig:borndiagram}
\end{figure}

At tree-level, the leading contribution to the process occurs via $t$ and $u$-channel charged Higgs-exchange diagram and the quartic coupling diagram. These contributions are at an order of $\mathcal{O}(\alpha_{ew})$ and they are based on pure electroweak interactions. The tree-level Feynman diagrams contributing to $\gamma\gamma \rightarrow H^{-}H^{+}$ in the 2HDM is given in Fig.~\ref{fig:borndiagram}.
The matrix elements corresponding to each diagram are respectively expressed as
\begin{equation}
\begin{split}
{\cal M}_1=&-iC_{H^{-}H^{+}\gamma \gamma} \text{g}^{\mu \nu} \varepsilon_{\mu}(p_1) \varepsilon_{\nu}(p_2),
\end{split}
\end{equation}
\begin{equation}
\begin{split}
{\cal M}_2=&\frac{ -iC_{H^{-}H^{+}\gamma}^2}{ \big[\hat{t}-m_{H^{\pm}}^2\big]} (p_2-2k_2)^\nu \varepsilon_{\nu}(p_2)\\
&\times(-p_2-k_1+k_2)^\mu \varepsilon_{\mu}(p_1),
\end{split}
\end{equation}
\begin{equation}
\begin{split}
{\cal M}_3=&\frac{-i C_{H^{-}H^{+}\gamma}^2}{ \big[\hat{u}-m_{H^{\pm}}^2\big]} (p_2-2k_1)^\nu \varepsilon_{\nu}(p_2)\\
&\times(-p_2+k_1-k_2)^\mu \varepsilon_{\mu}(p_1),
\label{eq:m1}
\end{split}
\end{equation}
where $\varepsilon_{\mu}(p_1)$ and $\varepsilon_{\nu}(p_2)$ are polarization vector of incoming photons. The total amplitude at the lowest order could be determined by summing up all the above
matrix elements:
\begin{equation}\label{eq:totMee}
{\cal M}_{0}=\sum_{i=1}^{3} {\cal M}_{i}.
\end{equation}
The tree-level total amplitude includes only the couplings
\begin{equation}\label{eq:cHHV}
\begin{split}
&C_{H^{-}H^{+}\gamma}=i e,\\
&C_{H^{-}H^{+}\gamma \gamma}=2 i e^2,
\end{split}
\end{equation}
which are independent of the 2HDM angles. These couplings are universal in a sense and come from the kinetic energy term of the Higgs fields. However, the other couplings, $C_{ h^0 H^- H^+}$ and $C_{ H^0 H^- H^+}$, are produced from the scalar potential of 2HDM. The couplings $C_{ h^0 H^- H^+}$ and $C_{ H^0 H^- H^+}$ mainly come up at $s$-channel diagrams.

After squaring the total amplitude and summing over the helicities of the final states, the total cross-section $\sigma(\gamma\gamma \rightarrow H^{-}H^{+})$ is obtained by using following formula
\begin{equation} \label{eq:totalsigma}
\hat{\sigma}_{\text{LO}}^{\gamma\gamma\rightarrow H^{-}H^{+}}=\frac
{1}{16\pi \hat{s}^{2}}\int_{\hat
t^{-}}^{\hat t^{+}} \biggl(\frac{1}{4}\biggr)\sum_{hel} |{\cal M}_0|^2 d\hat t
\end{equation}
with
\begin{equation}
\hat{t}^\pm=(m_{H^\pm}^2-\frac{\hat{s}}{2}) \pm\frac{1}{2}\bigl(\sqrt{\hat{s}^2-4 \hat{s} m_{H^\pm}^2 }\bigr).
\end{equation}
where the average over initial photons spins represented
by the factor (1/4). The parameters entering the tree-level cross-section are all standard except for the mass of the charged Higgs. Moreover, the non-standard parameters of 2HDM appearing at the one-loop level could be consistently taken as bare in the calculations.

\subsection{NLO Corrections}
In the analysis of high-energy processes observed at the current and future colliders, higher-order corrections (at least, next-to-leading order contributions) should be included  for precise theoretical predictions. The process~\eqref{eq:gammagammaHH} has one-loop level contributions as the next-to-leading order. They are based on pure EW interactions at the order of one-loop. In the one-loop level, the total amplitude can be expressed as a linear sum of triangle, box, and bubble one-loop integrals. According to the type of loop correction, the virtual contributions are coming from three different type diagrams: self-energy, box-type, and vertex-type (triangle$\&$bubble-type s-channel) diagrams.

\begin{figure*}[t]
    \begin{center}
\includegraphics[width=0.95\linewidth]{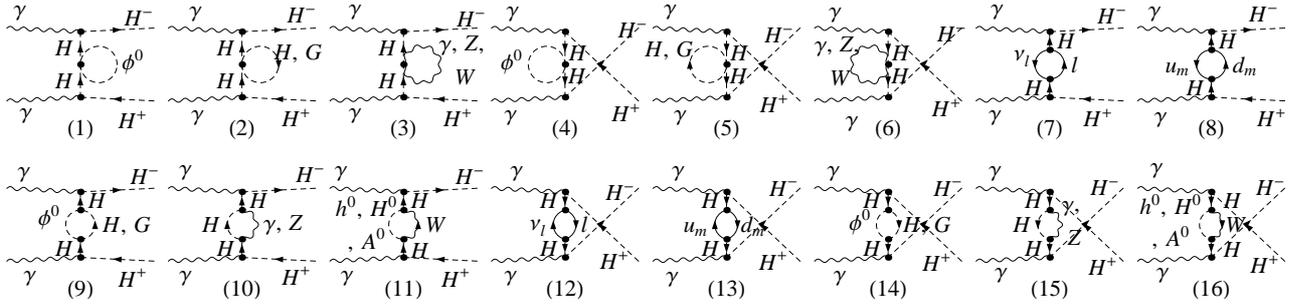}
     \end{center}
     \vspace{-4mm}
\caption{The self-energy correction diagrams contributing to the process $\gamma \gamma \rightarrow H^{-}H^{+}$.}\label{fig:diagself}
\end{figure*}

\begin{figure*}[t]
    \begin{center}
\includegraphics[width=0.95\linewidth]{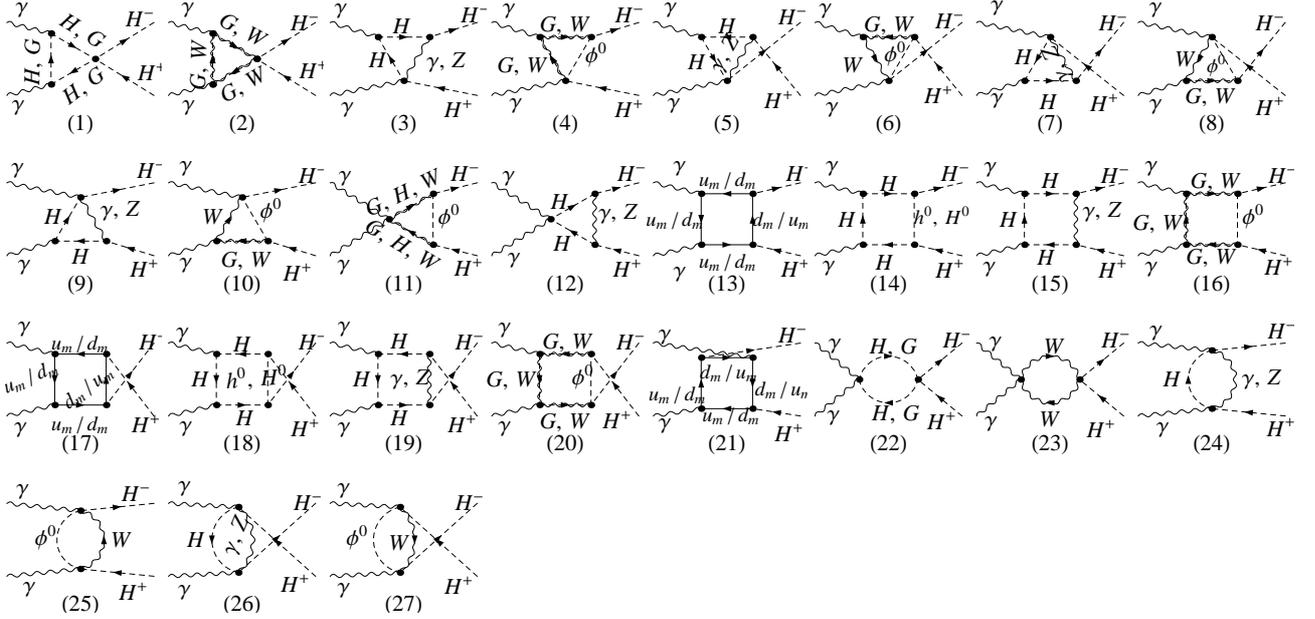}
     \end{center}
          \vspace{-4mm}
\caption{The box correction diagrams contributing to the process $\gamma \gamma \rightarrow H^{-}H^{+}$.}\label{fig:diagbox}
\end{figure*}

\begin{figure*}[t]
    \begin{center}
\includegraphics[width=0.95\linewidth]{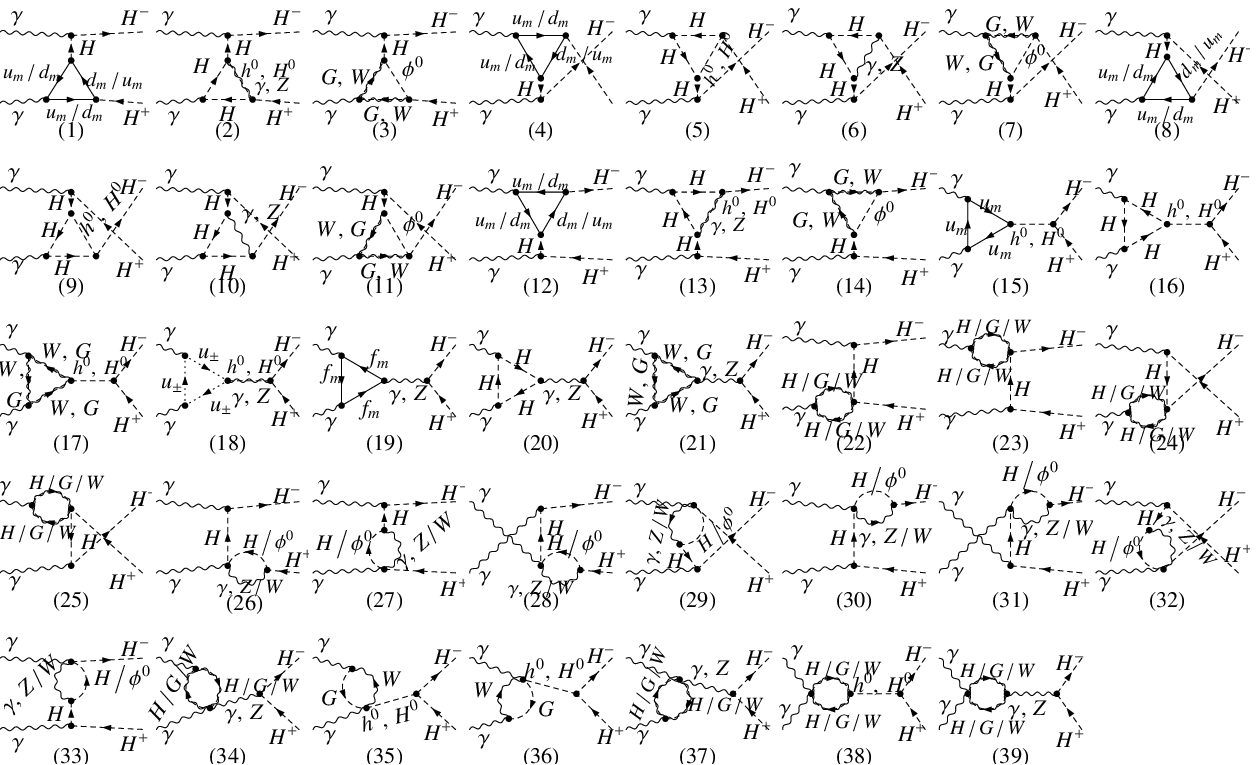}
     \end{center}
     \vspace{-4mm}
\caption{The vertex-correction diagrams contributing to the process $\gamma \gamma \rightarrow H^{-}H^{+}$.}\label{fig:diagvert}
\end{figure*}

A complete set of one-loop Feynman diagrams for $\gamma\gamma \rightarrow H^{-}H^{+}$ in the 2HDM and the corresponding amplitudes have been provided by the \textsc{FeynArts}. These diagrams are drawn in Figs.~\ref{fig:diagself} to~\ref{fig:diagvert}. In the other set of diagrams, particles in each loop are running counterclockwise. Note that Feynman diagrams with electron-Higgs couplings have been neglected. The internal particles in diagrams are labeled as follows: $\phi^0$ indicates to all neutral Higgs/Goldstone bosons ($h^0, H^0, A^0, G^0$); $u_{\pm}$ indicates the ghosts; $u_m/d_m$ can be $u/d$-type quarks (the subscript $m$ represents the generation of quark) and $l$ stands for leptons $e,\mu,\tau$. In loop diagrams, dashed-lines indicate to neutral and charged Higgs bosons, and wavy-lines represent $\gamma$ and $Z$, $W^\pm$-bosons. The Mandelstam variables~\eqref{eq:manvar} are also valid for in there. The contribution can also be topologically divided into $\hat{s}$, $\hat{t}$ and $\hat{u}$-channel diagrams with intermediate the neutral Higgs bosons ($h^0,H^0,A^0,G^0$), charged Higgs/Goldstone bosons ($H^\pm$,$G^\pm$), gauge bosons ($\gamma$, $Z$, $W^\pm$).

The self-energy diagrams consist of all possible self-energy loops of quarks, gauge bosons, and neutral/charged Higgs/Goldstone bosons on the propagator of charged Higgs boson as shown in Fig.~\ref{fig:diagself}. The box-type contributions include all possible loops of quarks, neutral Higgs bosons, gauge bosons, and charged Higgs/Goldstone bosons as shown in Fig.~\ref{fig:diagbox}.
The vertex-correction diagrams consist of triangle corrections to $\hat{t}$-channel charged Higgs exchange, triangle and bubbles vertices attached to the final state through an intermediate $\gamma$ or $Z$ or neutral Higgs bosons, as shown in Fig.~\ref{fig:diagvert}. Most of them consist mainly of $\hat{t}$ and $\hat{u}$-channel contributions. The $\hat{s}$-channel contributions only arrise from diagrams $(22)$-$(24)$ in Fig.~\ref{fig:diagbox} and diagrams $(15)$-$(21)$ and $(34)$-$(39)$ in Fig.~\ref{fig:diagvert}. The $\hat{s}$-channel diagrams may be make a significant contribution to the cross section, however they are nearly negligible away from the mass pole of the propagator.

\begin{figure}[hbt]
    \begin{center}
\includegraphics[width=1\linewidth]{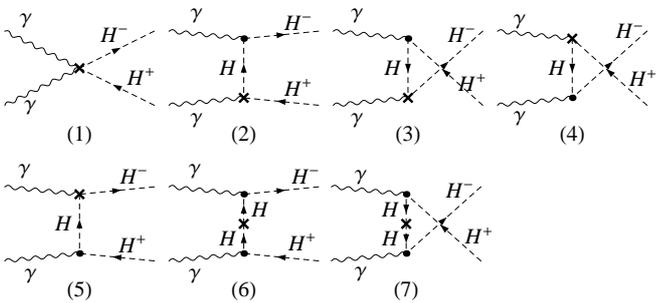}
     \end{center}
     \vspace{-4mm}
\caption{The counterterm diagrams for the process $\gamma\gamma \rightarrow H^{-}H^{+}$.}\label{fig:counter}
\end{figure}
\begin{figure}[t]
    \begin{center}
\includegraphics[width=1\linewidth]{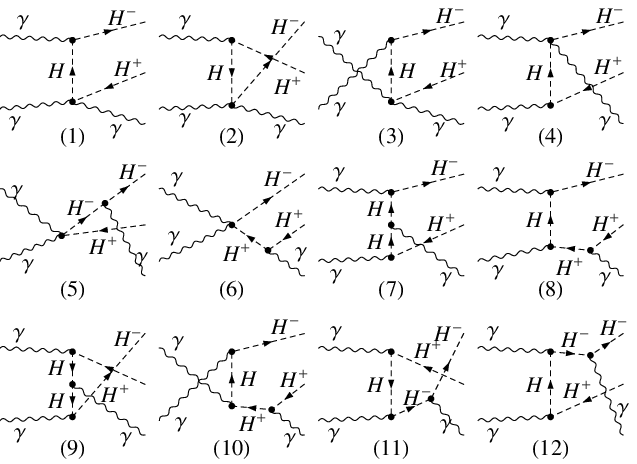}
     \end{center}
     \vspace{-4mm}
\caption{The Feynman diagrams for the real photon radiation.}\label{fig:radiation}
\end{figure}
The relevant total amplitude can be written by summation
over all contributions from self-energy, triangle,
and box diagrams as
\begin{equation}\label{eq:totalM}
{\cal M}_{\text{virt}}={\cal M}_{\bigcirc} + {\cal M}_{\Box}+  {\cal M}_{\triangle}.
\end{equation}
For virtual one-loop corrections, the differential cross section, summing over the helicities of the final states, can be calculated by
\begin{equation} \label{eq:dsigmavirt}
d\hat{\sigma}_{\text{virt}}^{\gamma\gamma\rightarrow H^{-}H^{+}}=\frac
{1}{16\pi \hat{s}^{2}} \biggl(\frac{1}{4}\biggr) \sum_{hel} 2 \text{Re}\bigl[{\cal M}_0^*{\cal M}_{\text{virt}}\bigr]d\hat{t}
\end{equation}
where $|{\cal M}_{\text{virt}}|^2$ is not included since it is very small.
The virtual contributions are ultraviolet (UV) and infrared (IR) divergent. The UV divergences are handled by dimensional regularization~\cite{Hooft72} in the on-mass-shell renormalization scheme. The counter terms are included via diagrams in Fig.~\ref{fig:counter}. For 2HDM, all Feynman rules including counter terms and the renormalization conditions are described in Ref.~\cite{Altenkamp17}. Here, the calculation has been carried out by using the \textsc{FeynArts} model file that includes these counter terms (for details see Ref.~\cite{Altenkamp17}). The calculation of the amplitude has been performed in 't Hooft-Feynman gauge. After the renormalization procedure, the virtual part becomes UV-finite. Though, it still contains the soft IR singularities originated from the exchange of virtual photons in the loops. These singularities are regularized by introducing a photon mass parameter, $m_\gamma$\footnote{This is automatically carried out by \textsc{LoopTools}.}. All these singularities in the limit $m_\gamma \rightarrow 0$ are cancelled by adding the real photon bremsstrahlung corrections, according to the Kinoshita-Lee-Nauenberg theorem~\cite{Kinoshita62,Lee64}. The real photon radiation process is denoted by
\begin{equation} \label{eq:gammagammaHHgamma}
\gamma(p_1)\gamma(p_2)\rightarrow H^{-}(k_1) H^{+}(k_2) \gamma(k_3),
\end{equation}
where $k_3$ is the four-momenta of radiated photon. The relevant diagrams are given in Fig.~\ref{fig:radiation}.
According to the energy of the radiation photon $k_3^0=\sqrt{|\overrightarrow{k_3}|^2+m_\gamma^2}$, the bremsstrahlung phase space is divided into a soft region and a hard region. Hence, the real photon radiation correction is written as follows:
\begin{equation} \label{eq:dsigmaSB}
d\hat{\sigma}_{\text{real}}^{\gamma\gamma\rightarrow H^{-}H^{+}\gamma}=d\hat{\sigma}_{\text{soft}}(\delta_s)+d\hat{\sigma}_{\text{hard}}(\delta_s)
\end{equation}
where $\delta_s$ is the soft cut-off energy parameter $\delta_s=\Delta E_\gamma/(\sqrt{\hat{s}}/2)$. If the energy of radiation photon is $k_3^0<\Delta E_\gamma=\delta_s \sqrt{\hat{s}}/2 $, it is called soft. If $k_3^0>\Delta E_\gamma$, the radiation photon is hard. The soft part is calculated by using the soft photon approximation
formula~\cite{Hooft79,Denner93}
\begin{equation} \label{eq:dsoft}
\begin{split}
d\hat{\sigma}_{\text{soft}}=-d\hat{\sigma}_{0}\frac{\alpha e^2}{2\pi^2} \int_{|\overrightarrow{k_3}|\leq \Delta E_\gamma} \frac{d^3k_3}{2k^0_3} \biggl[\frac{k_1}{k_1\cdot k_3}-\frac{k_2}{k_2\cdot k_3}\biggl]^2
\end{split}
\end{equation}
where $d\hat{\sigma}_{0}$ is the tree-level differential cross section and the soft photon cut-off energy $\Delta E_\gamma$ satisfies $k^0_3\leq\Delta E_\gamma \ll\sqrt{\hat{s}}$.

Although both soft and hard terms depend on soft cut-off parameter $\delta_s$, the real correction does not depend on the soft cut-off parameter. Furthermore, summing the virtual and soft contributions drops out the dependence of the IR regulator $m_\gamma$. The result now depends on the parameter $\delta_s$, i.e., $\Delta E_\gamma$, and the hard photon radiation contribution must be added as well for dropping this dependence out.

Consequently, the UV and IR finite total one-loop corrections are expressed as a sum of the virtual, the soft photon radiation, and the hard photon radiation:
\begin{equation} \label{eq:dsigmaNLO}
\begin{split}
d\hat{\sigma}_{\text{NLO}}^{\gamma\gamma\rightarrow H^{-}H^{+}}&=d\hat{\sigma}_{\text{virt}}(m_\gamma)+d\hat{\sigma}_{\text{soft}}(m_\gamma,\delta_s)\\
&+d\hat{\sigma}_{\text{hard}}(\delta_s)
\end{split}
\end{equation}
which is independent of the IR regulator $m_\gamma$ and soft cut-off parameter $\delta_s$.

\begin{figure}[!hbt]
    \begin{center}
\includegraphics[scale=0.42]{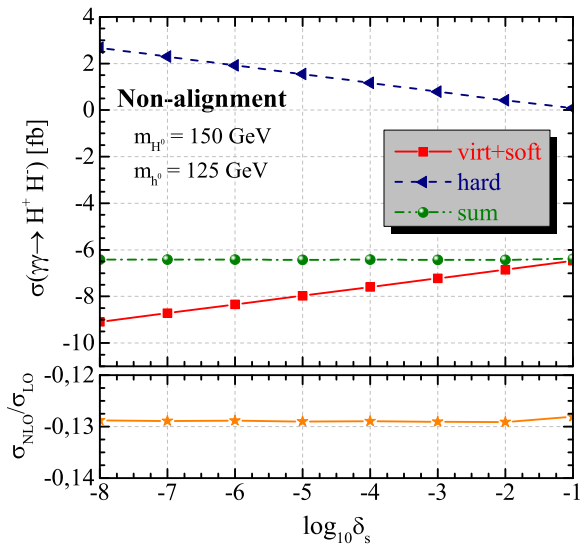}
     \end{center}
     \vspace{ -5mm}
\caption{(color online). The virtual, soft and hard photon radiation corrections to process $\gamma\gamma \to H^- H^+  $ as a function of the soft cutoff $\delta_s$ for the non-alignment scenario.}
\label{fig:delts}
\end{figure}

We have numerically checked that our results do not depend on $m_\gamma$ or on $\Delta E_\gamma=\delta_s \sqrt{\hat{s}}/2 $. For the representatively non-alignment scenario with $t_\beta=10$ and $m_{H^0}=150\gev$,  the virtual plus soft correction, the hard photon radiation correction and the total one-loop correction are plotted as a function of the soft cutoff parameter $\delta_s$ at $\sqrt{\hat{s}}=1\tev$ in Fig.~\ref{fig:delts}.
As one can see from this figure, the virtual plus soft correction and the hard photon radiation correction change with the variation of $\delta_s$, but their sum remains almost constant, i.e., does not change over several orders of magnitude. The relative one-loop correction $\hat{\sigma}_{\text{NLO}}/\hat{\sigma}_{\text{LO}}$ is also stable around $-13\%$ as shown in subpanel of figure.

During our numerical evaluation below, the soft cutoff parameter has been fixed as $\delta_s = 10^{-3}$.

\subsection{Calculation of the Parent Process $e^+e^-\rightarrow\gamma\gamma\rightarrow  H^+ H^-$}
The photon-photon collisions can be realized at the facility in the future generation of TeV-class linear colliders. Then $\gamma\gamma\to H^+ H^-$ is generated as a subprocess of electron-positron collision at the linear colliders. The total cross-section of the parent process $e^+e^-\rightarrow\gamma\gamma\rightarrow H^+ H^-$ could be obtained by folding $\hat{\sigma}(\gamma\gamma\rightarrow H^+ H^-)$ with the luminosity of photon
\begin{equation}
\frac{dL_{\gamma\gamma}}{dz}=2z\int_{z^2/x_{max}}^{x_{max}}\frac{dx}{x}F_{\gamma/e}(x)F_{\gamma/e}\left(\frac{z^2}{x}\right)\,,
\end{equation}
so that
\begin{align} \label{eq:total_cross}
\begin{split}
&\sigma^{e^+e^-\rightarrow\gamma\gamma\rightarrow H^+ H^-}(s)=\\
&\int_{(2m_{H^\pm})/\sqrt{s}}^{x_{max}} dz \frac{dL_{\gamma\gamma}}{dz}~ \hat{\sigma}( \gamma\gamma\rightarrow  H^+ H^-;\; \hat{s}=z^2s ),
\end{split}
\end{align}
where $F_{\gamma/e}(x)$ represents the photon structure function. The photon spectrum is qualitatively better for larger values of the $x$-fraction of the longitudinal momentum of the $e^-$-beam. In the case of $x > 2(1+\sqrt{2})\approx 4.8$, the high-energy photons could vanish via the pair production of $e^-e^+$ in its collision with a subsequent laser-$\gamma$. The energy spectrum of the photon provided as a Compton backscattered photon off the $e^-$-beam~\cite{Telnov} has been utilized for $F_{\gamma/e}(x)$ in this study.

\section{Numerical Results And Discussions} \label{sec:results}
The numerical results of the production of the charged Higgs boson pairs through photon-photon collisions are discussed in detail, considering full one-loop corrections in the 2HDM, including soft and hard QED radiation. For each benchmark scenario, the tree-level and the NLO corrections of the cross sections are numerically evaluated as a function of the center-of-mass energy and the mass of Higgs boson selected as a non-fixed free parameter. The regions of the parameter space where the production rates are large enough to be detectable are highlighted. The longitudinal polarizations of the initial beams are significant to improve the production rate; therefore, some polarization distributions are presented, as well. Decay channels of the charged Higgs boson are also investigated for the scenarios interested.

For the numerical calculation, the SM input parameters are set as $G_F = 1.1663787(6)\times10^{-5}\gev^{-2}$, $m_W= 80.385\gev$, $m_Z= 91.1876\gev$,  $m_t=173.21\gev$, and $\alpha^{-1}(0)= 137.03599$.

The following notations are used here:
\begin{itemize}
  \item[i.] $\sigma_{\text{LO}}:=$  the tree-level total cross section.
  \item[ii.] $\sigma_{\text{NLO}} :=$   the NLO corrections of the cross-section i.e., the virtual plus real contributions.
  \item[iii.] $\sigma_{\text{LO+NLO}} :=$  the full cross section including all one-loop corrections.
  \item[iv.] $\delta_r :=$               the relative correction in percent defined as $(\sigma_{\text{NLO}}/\sigma_{\text{LO}})\times 100$.
\end{itemize}

In the following subsections, the numerical evaluations for each BP is separately presented. The numerical results are of course dependent on the choice of the 2HDM parameters. Nonetheless, they provide an opinion of the relevance of the full one-loop contributions. As a general comment, it can be pointed out that the tree-level cross sections depend solely on the parameters of SM (and $m_{H^\pm}$). As a result, any dependence on the free parameters of 2HDM can appear firstly at the one-loop level (except for $m_{H^\pm}$).

\subsection{Non-alignment scenario}
\begin{figure*}[!hbt]
    \begin{center}
\includegraphics[scale=0.38]{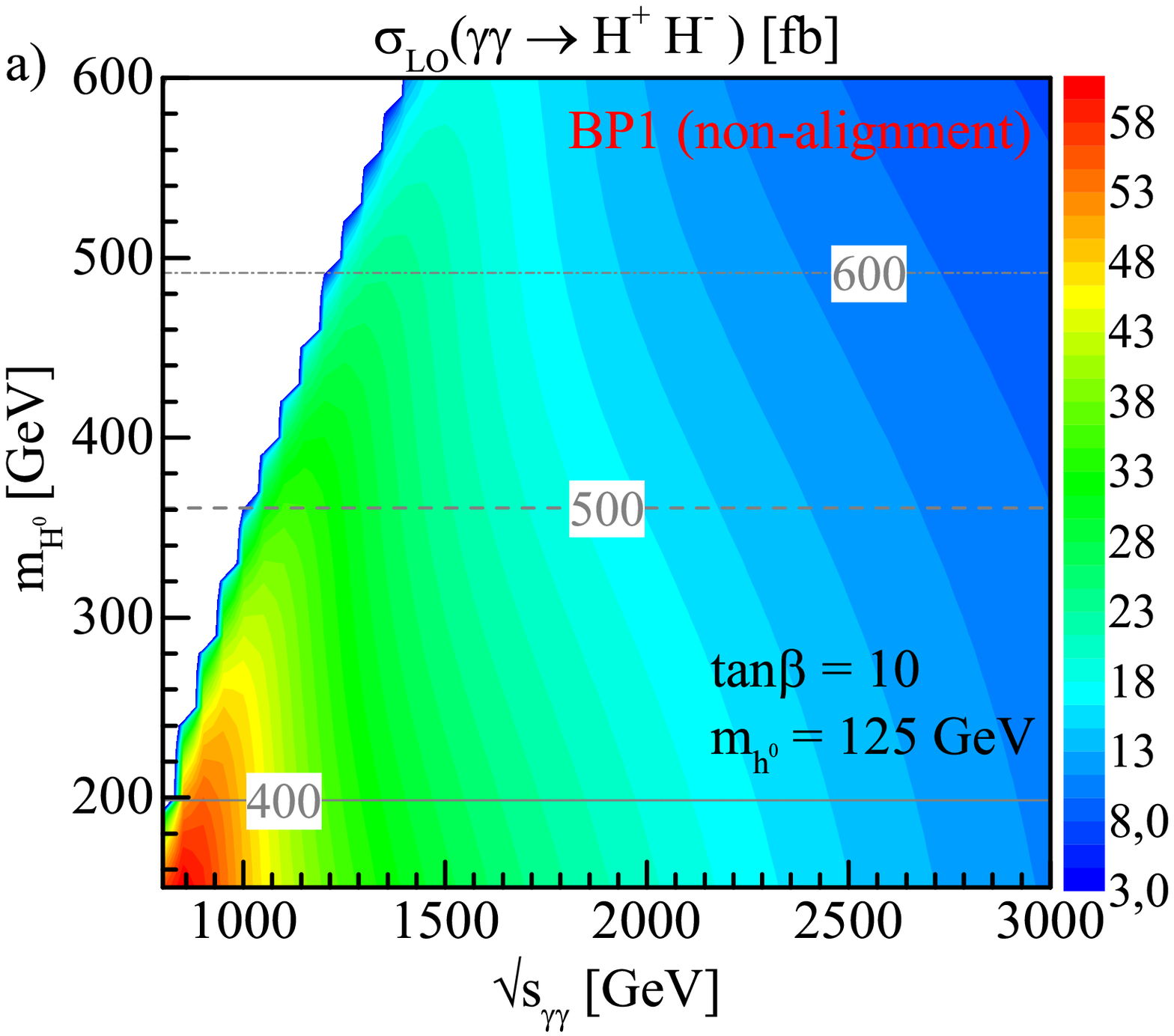}
\includegraphics[scale=0.38]{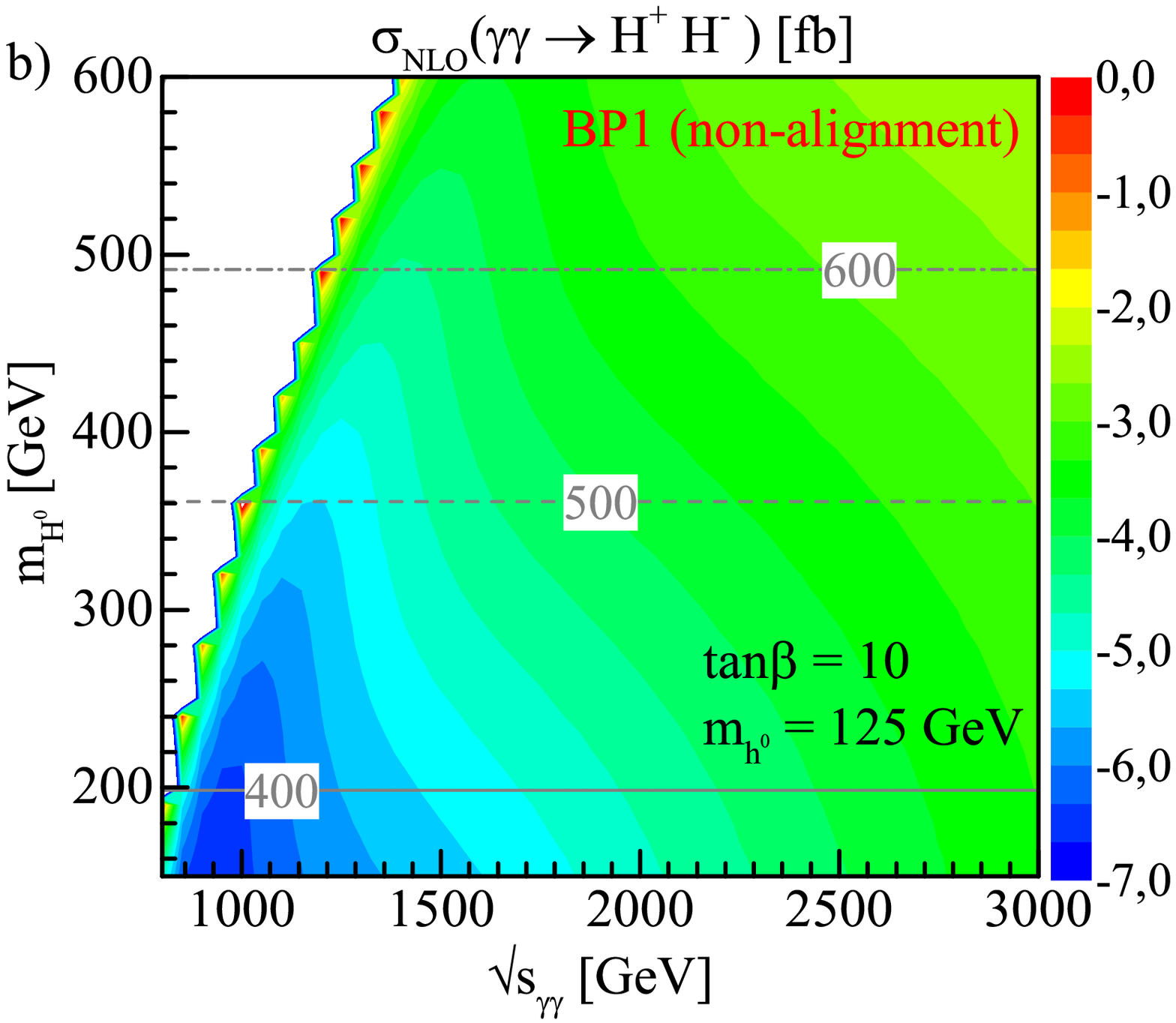}
\includegraphics[scale=0.38]{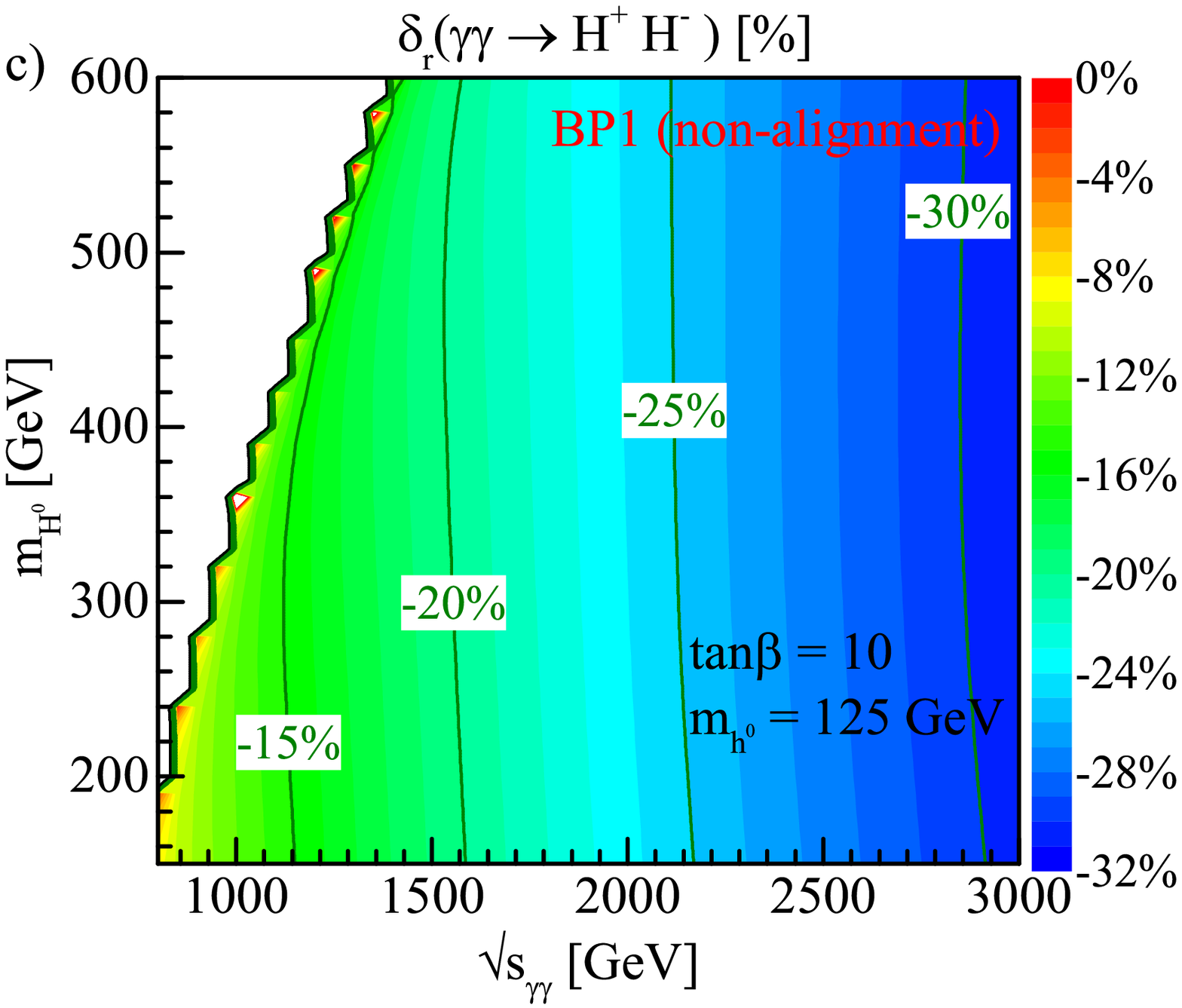}
     \end{center}
     \vspace{ -6mm}
\caption{(color online). (a) The tree-level and (b) NLO corrections of cross sections (in fb), and (c) the corresponding relative correction for process $ \gamma \gamma \to H^- H^+$ scanned over the ($m_{H^0}$, $\sqrt{s}_{\gamma\gamma}$) plane with $\tan \beta=10$ in the non-alignment scenario. The contour lines correspond to predictions for $m_{H^\pm}$ in unit of GeV, and the relative corrections as a percentage.}
\label{fig:BP1}
\end{figure*}
For BP1 defined in the non-alignment scenario, the tree-level and the NLO corrections of cross sections of process $\gamma\gamma \to H^- H^+ $ are scanned over the regions of $m_{H^0}$-$\sqrt{s}$, and plotted in Figs.~\ref{fig:BP1}(a)-(b), respectively. Additionally, to describe the full EW corrections to the tree-level cross section quantitatively, the corresponding relative correction is plotted as functions of $m_{H^0}$-$\sqrt{s}$ in Fig.~\ref{fig:BP1}(c). The scan parameters are varied as follows: $150 \leq m_{H^0} \leq 600\gev$ in steps of 10 GeV, and $850 \leq \sqrt{s} \leq  3000\gev$ in steps of 50 GeV. Here, it is assumed that the lightest CP-even scalar is the SM-like Higgs boson. For this scenario, the vacuum stability and perturbativity do not allow a large split between $m_{A^0}$ and $m_{H^\pm}$; for this reason it is set $m_{A^0}=m_{H^\pm}$. Also, any value of $m_{A^0}$
is allowed via the EW precision tests. Consequently, this scenario have a mass hierarchy as follow: $m_{h^0}=125\gev<m_{H^0}<m_{H^\pm}=m_{A^0}$. The values of the mass of charged Higgs boson $m_{H^\pm}$, which are calculated by Eq.~\eqref{eq:mcHmA} in terms of the mass of neutral Higgs boson $m_{H^0}$, are shown with contour lines in Figs.~\ref{fig:BP1}(a)-(b), and the $m_{H^\pm}$ grows relatively with increasing of the $m_{H^0}$. When $m_{H^0}$ runs from 150 to 600 GeV, $m_{H^\pm}$ varies from $379$ to $691\gev$ for BP1.

We can see from these figures that the total cross sections are sensitive to the mass of Higgs boson $m_{H^0}$ since $m_{H^0}$ is directly related with $m_{H^\pm}$ so the phase space of the final state particles. In the case of $m_{H^\pm} > \sqrt{s}/2$, the charged Higgs pair production is kinematically inaccessible as shown by white regions in parameter space. When $m_{H^\pm} \ll \sqrt{s}/2$, the cross section nearly scales as $1/s$ and reaches its larger values. When the mass of charged Higgs boson $m_{H^\pm}$ with along $m_{H^0}$ becomes larger, the tree-level cross section decreases as expected while NLO corrections slowly increases. Particularly, the cross section reaches its larger values for $m_{H^0}< 400\gev$ in the scan region. The NLO corrections make negative contributions to total cross section except for extreme points in the parameter regions. The NLO correction of total cross section $\sigma_{\text{NLO}} (\gamma\gamma \rightarrow H^{-}H^{+})$ ranges from -7 to -2 fb at most of the parameter space. The relative correction is always negative in the whole region and mostly ranges from $-15\%$ to $-30\%$ as seen from the contour lines in Fig.~\ref{fig:BP1}(c). Its magnitude increases with increasing of $\sqrt{s}_{\gamma\gamma}$ and reaches about $-31\%$ at $\sqrt{s}_{\gamma\gamma}=3\tev$. However, the relative correction becomes larger near the production threshold ($\sqrt{s}_{\gamma\gamma}\thickapprox 2 m_{H^\pm}$) because the cross section is very small in this region, so this enhancement is phenomenologically insignificant. At $\sqrt{s}_{\gamma\gamma}=1\tev$, the relative correction is about $-13.06\%$ for $m_{H^\pm}=401.25\gev$ and $-8.38\%$ for $m_{H^\pm}=492.63\gev$. For example, at $\sqrt{s}_{\gamma\gamma}=900\gev$ for $m_{H^0}=150\gev$ ($m_{H^\pm}=379.05\gev$), the full one-loop corrected cross section is $\sigma_{\text{LO+NLO}} (\gamma\gamma \rightarrow H^{-}H^{+})=51.97$ fb with $\delta_r =-11.28\%$. Overall, the full one-loop corrected cross section is at a visible level of $\mathcal{O}( 10^1~\text{fb})$ in the range of 4 to 55 fb for considered parameter regions of non-alignment-scenario.

\begin{figure}[!hbt]
    \begin{center}
\includegraphics[scale=0.39]{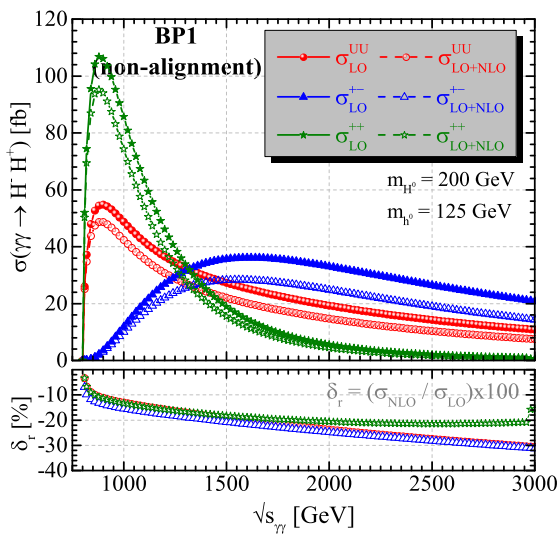}
     \end{center}
     \vspace{ -6mm}
\caption{(color online). The polarized tree-level and full one-loop EW-corrected cross sections of process
$\gamma\gamma \to H^- H^+  $ for different polarization modes
of initial beams as a function of $\sqrt{s}_{\gamma \gamma}$ with $m_{H^0}=200\gev$ and $\tan \beta=10$ for BP1.
}
\label{fig:BP1pol}
\end{figure}

In Fig.~\ref{fig:BP1pol}, the initial beam polarisation dependence of the integrated tree-level and full one-loop EW-corrected cross sections are plotted as a function of $\sqrt{s}_{\gamma\gamma}$ for BP1, where we take $m_{H^0}=200\gev$ and $t_\beta=10$. The $\sqrt{s}_{\gamma\gamma}$ varies from the value little larger than the threshold $2 m_{H^\pm}$ to $3\tev$. The curves correspond to the integrated cross section with oppositely-polarized photons $(+-)$, right-handed polarized photons $(++)$ and unpolarized photons $(\text{UU})$, respectively. Note that the integrated cross sections with the $(+-)$ and $(-+)$ photon polarization are equal: $\sigma^{+-}=\sigma^{-+}$.
As indicated in this figure, all curves reach to their maximum cross section values as the center-of-mass energy goes from the threshold value to the corresponding position of peak. This is also the expected behavior. These peaks appear at $\sqrt{s}_{\gamma\gamma}=890\gev$ for $\sigma^{\text{UU}}$, $\sqrt{s}_{\gamma\gamma}=880\gev$ for $\sigma^{++}$ and $\sqrt{s}_{\gamma\gamma}=1600\gev$ for $\sigma^{+-}$.  At high energies, the integrated cross sections with oppositely polarized photons $(-+)$ or $(+-)$ are enhanced by a factor of 1.9 as compared to the unpolarized case. In the case of both photons with right-handed or left-handed polarized $(++)$ or $(--)$, although at high energies, the integrated cross sections are highly suppressed, at low energies they are amplified up to about 2 times. These results imply that having both photons polarized can turn out to be significant to ensure a measurable production rate. The $\sigma^{\text{UU}}_{\text{LO+NLO}}$ reaches a maximum of 48.78 fb and the corresponding relative correction $\delta^{UU}_r$ is $-10.94\%$. The $\sigma^{++}_{\text{LO+NLO}}$ reaches a maximum of 95.21 fb and the corresponding relative correction $\delta^{++}_r$ is $-10.82\%$. The $\sigma^{+-}_{\text{LO+NLO}}$ reaches a maximum of 28.3 fb and the corresponding $\delta^{+-}_r$ is $-21.66\%$. When the $\sqrt{s}_{\gamma\gamma}$ goes from $1$ to $1.5\tev$, the unpolarized cross section $\sigma^{\text{UU}}_{\text{LO+NLO}}$ decreases from 42.39 to 22.04 fb. In the same energy range, the $\sigma^{++}_{\text{LO+NLO}}$ decreases from 74.27 to 15.49 fb.

On the other hand, the absolute relative corrections increase with the increment of $\sqrt{s}_{\gamma\gamma}$ for all cases. The relative corrections $\delta_r$ change in the ranges of $\delta^{\text{UU}}_r \in [-3.4\%, -30.6\%]$,  $\delta^{+-}_r \in [-6.9\%, -31.3\%]$ and  $\delta^{++}_r \in [-3.4\%, -15.7\%]$, when the $\sqrt{s}_{\gamma\gamma}$ goes from $810\gev$ to $3\tev$.

\subsection{Short-cascade scenario}
\begin{figure*}[!hbt]
    \begin{center}
\includegraphics[scale=0.39]{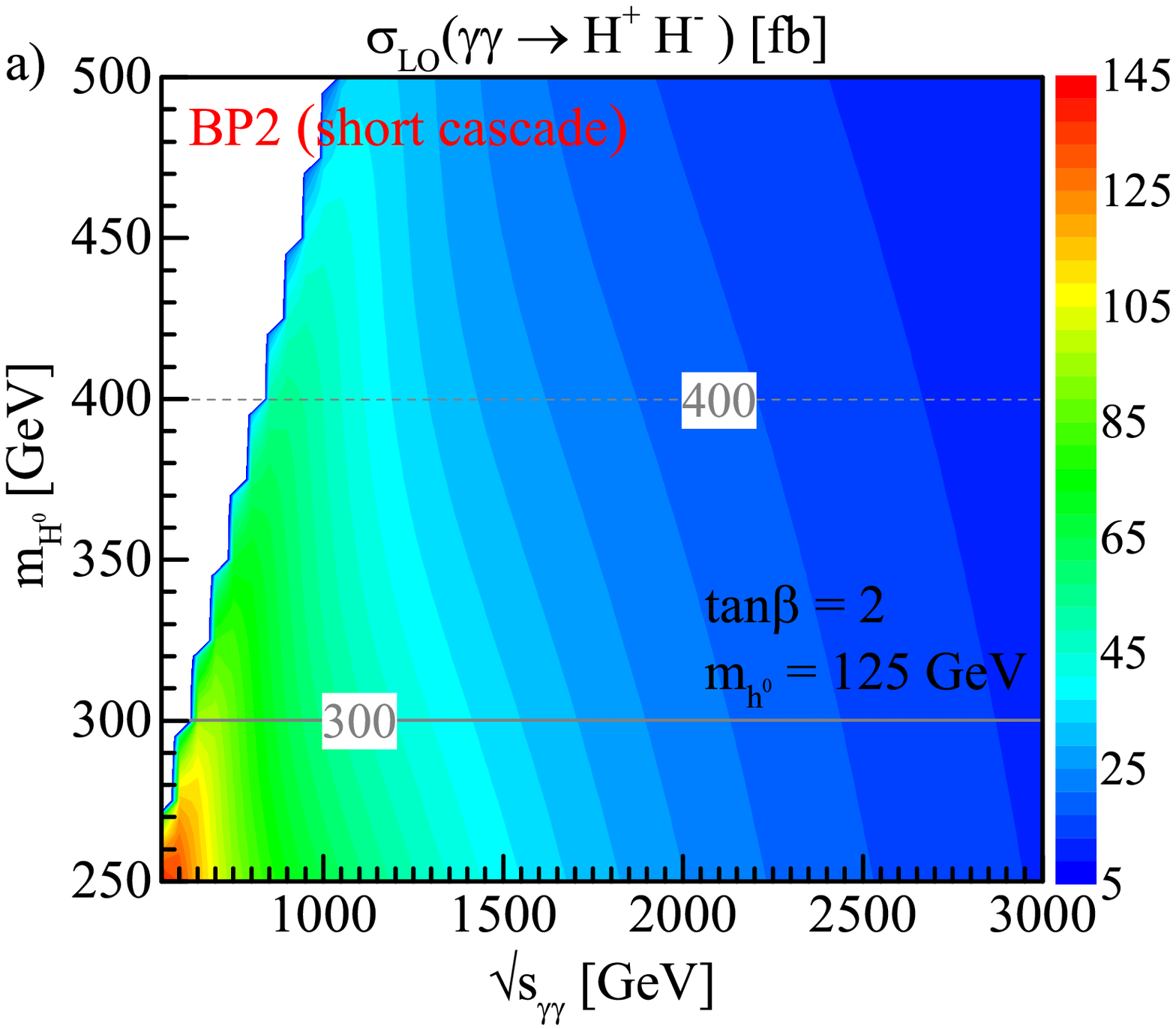}
\includegraphics[scale=0.39]{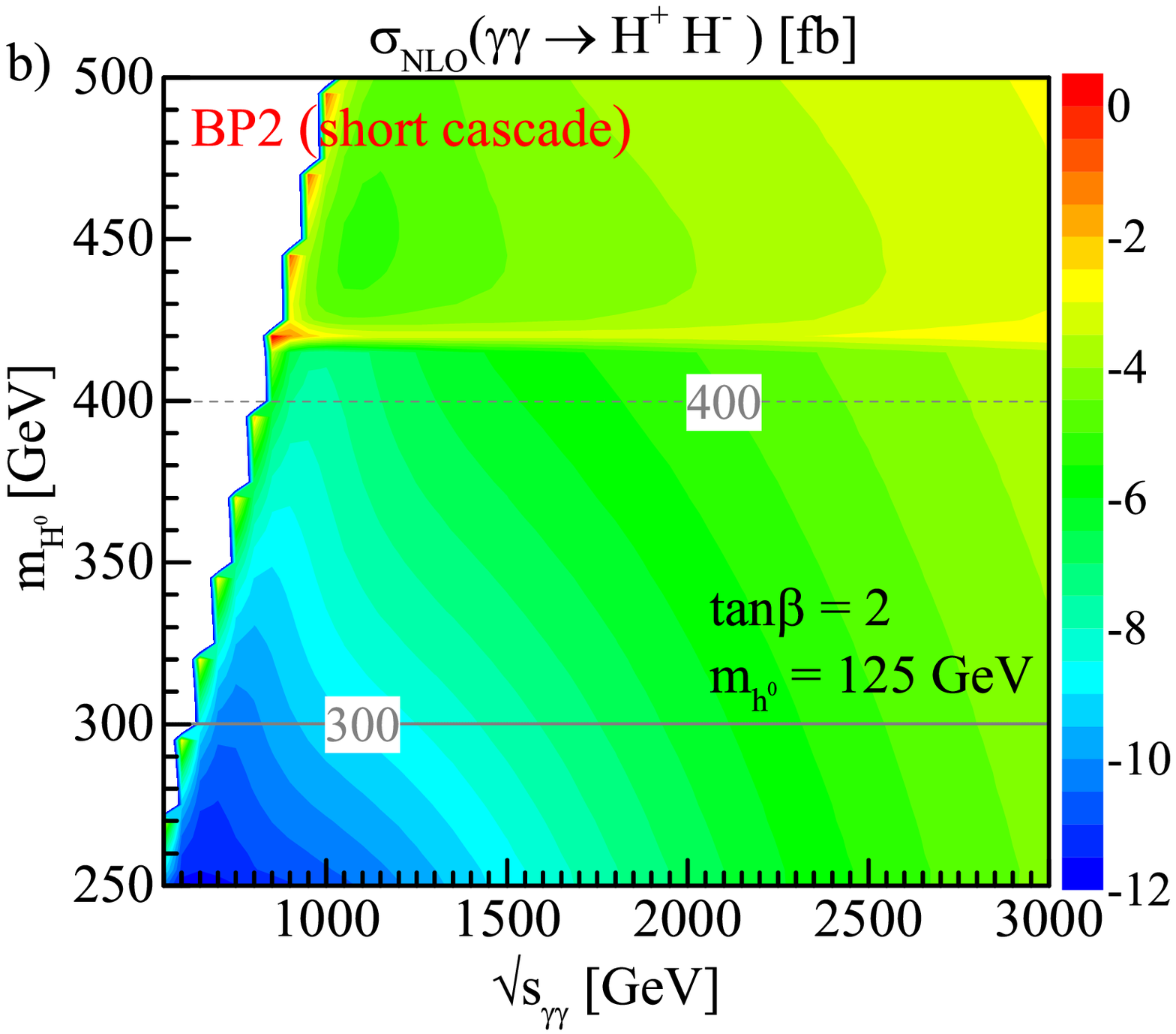}
\includegraphics[scale=0.39]{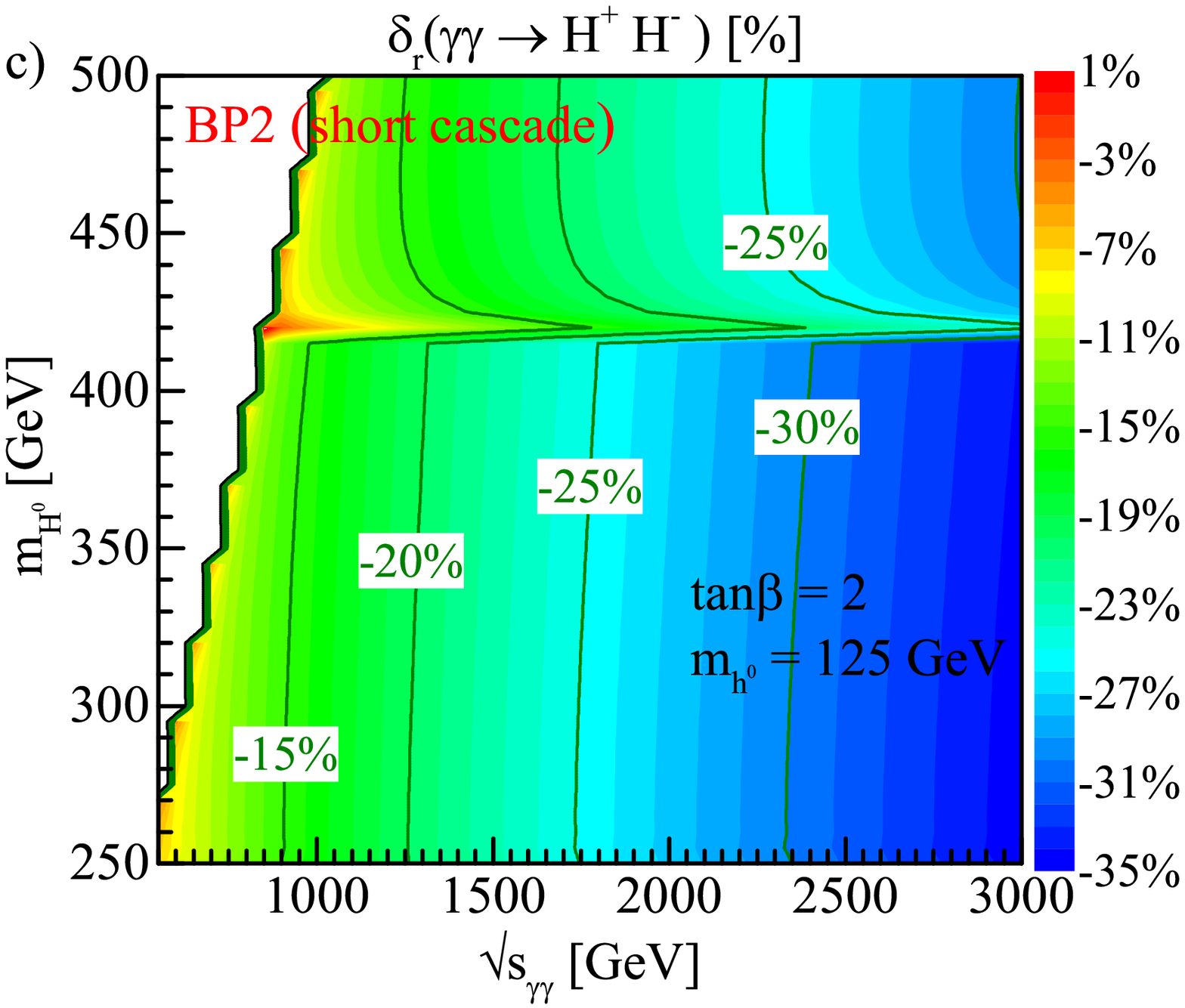}
     \end{center}
     \vspace{ -6mm}
\caption{(color online). (a) The tree-level and (b) NLO corrections of cross sections (in fb), and (c) the corresponding relative correction for process $ \gamma \gamma \to H^- H^+$ scanned over the ($m_{H^0}$, $\sqrt{s}_{\gamma\gamma}$) plane for BP2 in the short-cascade scenario. The contour lines correspond to predictions for $m_{H^\pm}$ in unit of GeV, and the relative corrections as a percentage.}
\label{fig:BP2}
\end{figure*}

\begin{figure*}[!hbt]
    \begin{center}
\includegraphics[scale=0.39]{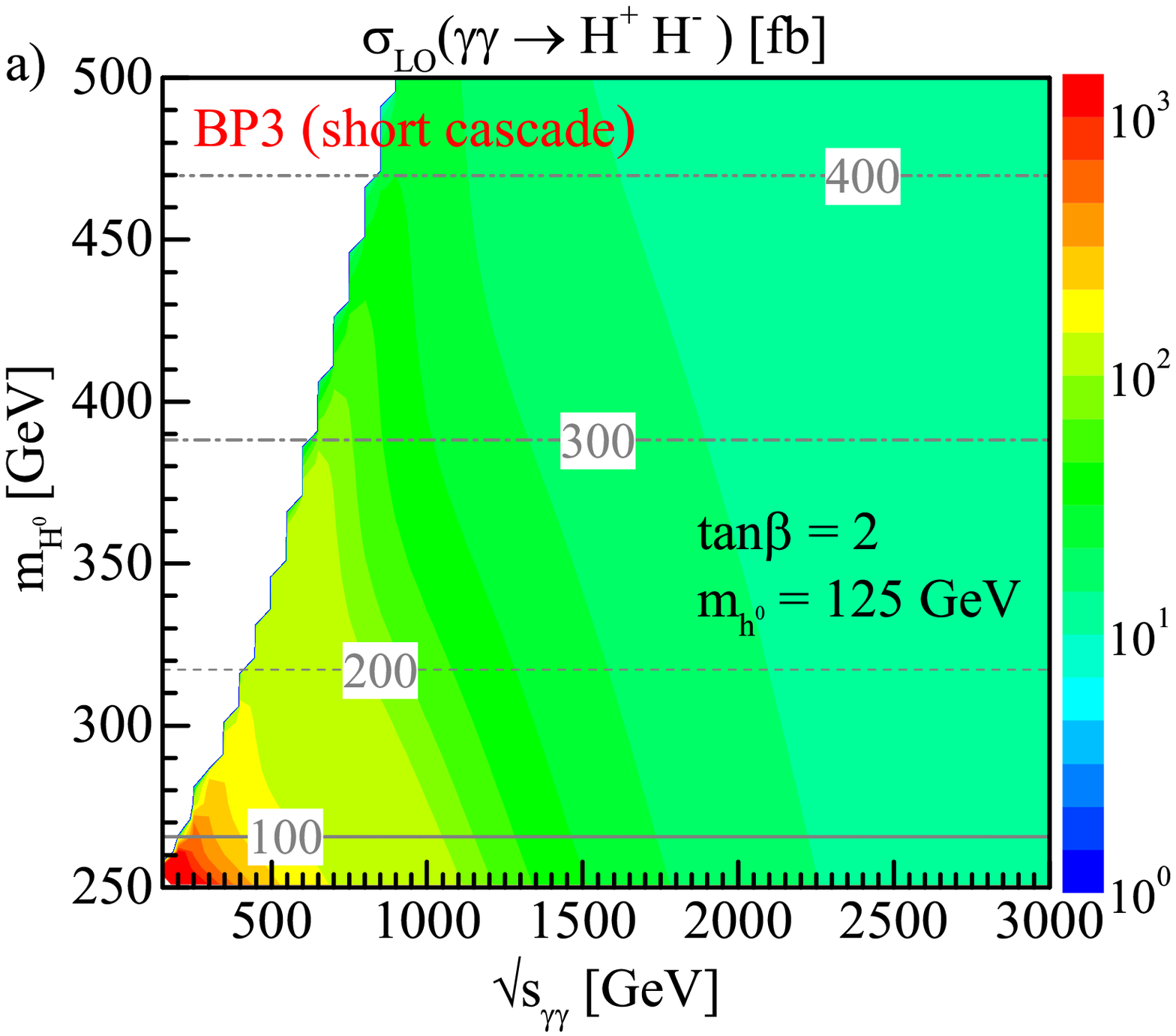}
\includegraphics[scale=0.39]{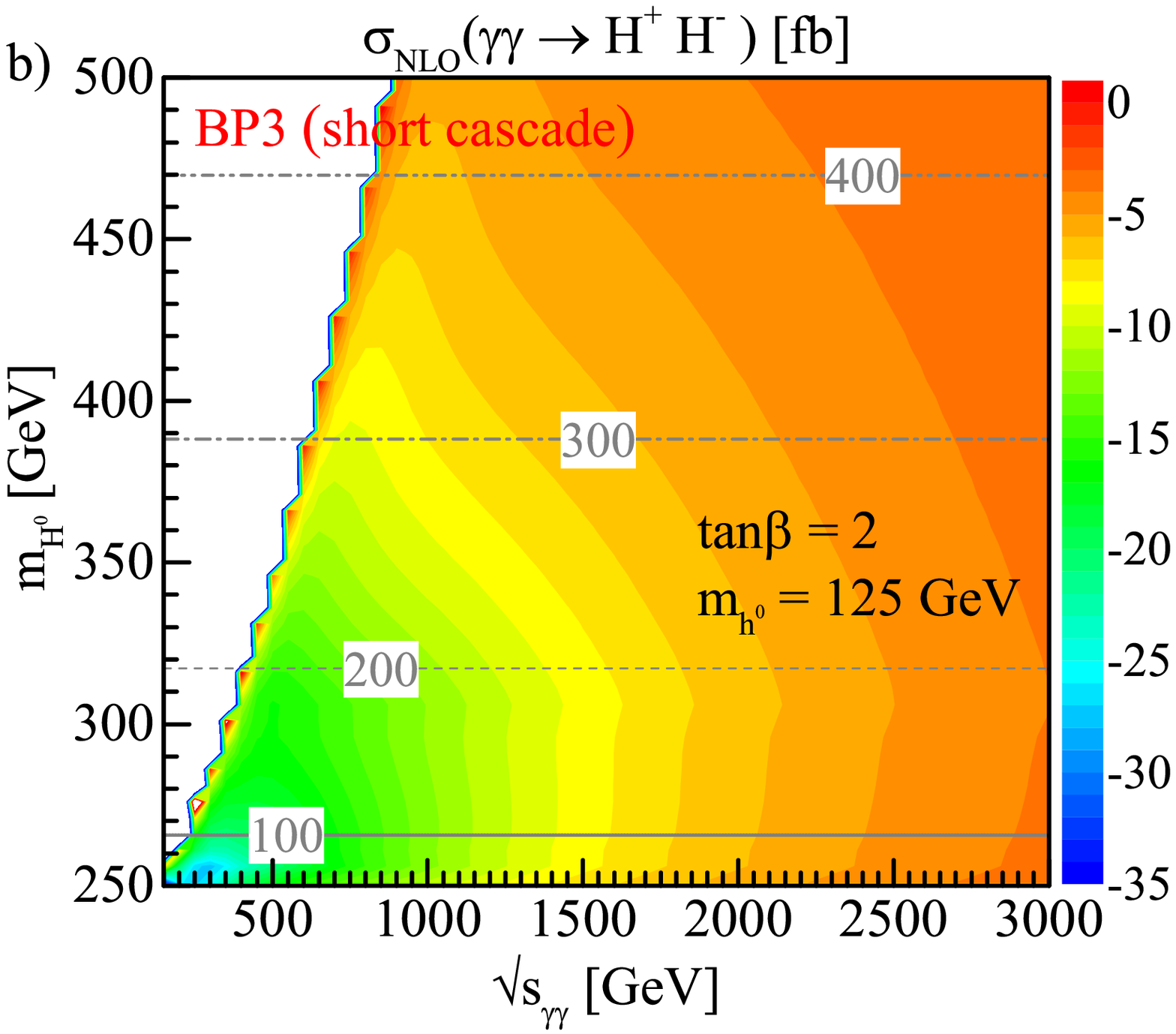}
\includegraphics[scale=0.39]{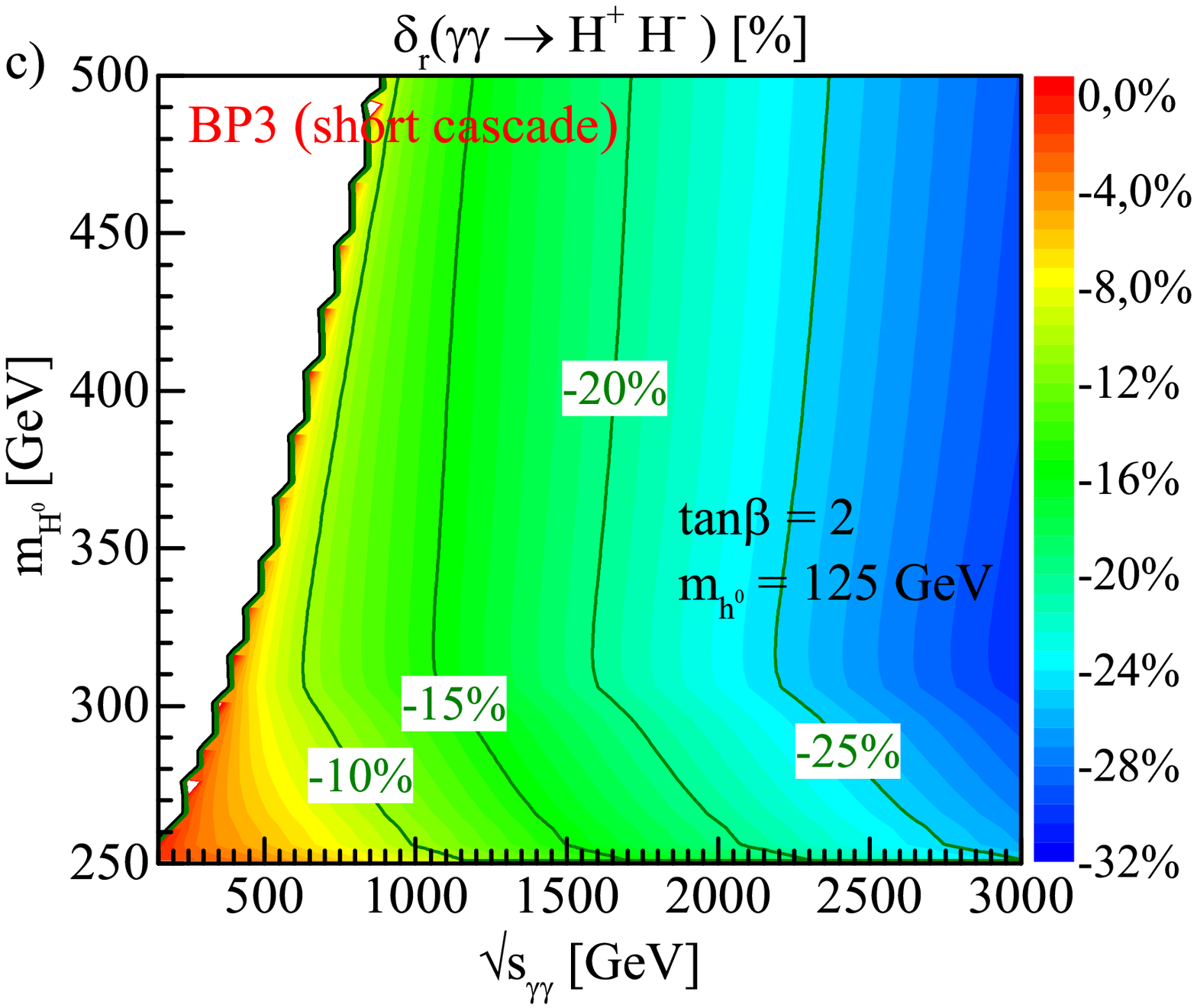}
     \end{center}
     \vspace{ -6mm}
\caption{(color online). (a) The tree-level and (b) NLO corrections of cross sections (in fb), and (c) the corresponding relative correction for process $ \gamma \gamma \to H^- H^+$ scanned over the ($m_{H^0}$, $\sqrt{s}_{\gamma\gamma}$) plane for BP3 in the short-cascade scenario. The contour lines correspond to predictions for $m_{H^\pm}$ in unit of GeV, and the relative corrections as a percentage.}
\label{fig:BP3}
\end{figure*}

The tree-level and the NLO corrections of cross sections, and the corresponding relative correction of process $\gamma\gamma \to H^- H^+ $ are scanned over plane of $m_{H^0}$-$\sqrt{s}_{\gamma\gamma}$ for BP2 (Fig.~\ref{fig:BP2}) and BP3 (Fig.~\ref{fig:BP3}) in the short-cascade scenario, where the lightest CP-even Higgs $h^0$ behaves like the SM Higgs boson by fixing $c_{\beta-\alpha}$ to be zero (exact alignment). The scan parameters are varied as follows: $250 \leq m_{H^0} \leq 500\gev$ in steps of 5 GeV, and $150-550 \leq \sqrt{s} \leq  3000\gev$ in steps of 50 GeV. There are two mass hierarchies as follows: $m_{A^0}<m_{H^0}=m_{H^\pm}$ for BP2 and $m_{H^\pm}<m_{A^0}=m_{H^0}$ for BP3. The values of the mass of charged Higgs boson $m_{H^\pm}$, which are calculated by Eq.~\eqref{eq:mcHmA} in terms of the mass of neutral Higgs boson $m_{H^0}$, are shown with contour lines, and the $m_{H^\pm}$ increases with increasing of the $m_{H^0}$.  The white regions in parameter space mean that the charged Higgs pair production is kinematically inaccessible. When $m_{H^\pm} \ll \sqrt{s}/2$, the cross section approximately scales as $1/s$, the $t$-channel contributions to production rate become important. When $m_{H^0}$ runs from 250 to 500 GeV, $m_{H^\pm}$ varies from $250$ to $500\gev$ for BP2, while it varies from $48.75$ to $436\gev$ for BP3. Due to presence of exact alignment $c_{\beta-\alpha}=0$ in the short-cascade scenario, the coupling $C_{h^0 H^- H^+}$ would reach its largest value affecting the cross section. However, the effect of $C_{H^0 H^- H^+}$ on the total cross-section will be reduced.

For both benchmark points, the NLO corrections make negative contributions to total cross section except for threshold points in the parameter space. The tree-level cross section decreases with increment of $m_{H^\pm}$ as expected, while the NLO corrections increases quantitatively. The size of NLO-corrected cross sections reach up to a level of $10^{2}$ and $10^{3}$ fb for BP2 and BP3, respectively.

As shown in Fig.~\ref{fig:BP2}, the $\sigma_{\text{NLO}} (\gamma\gamma \rightarrow H^{-}H^{+})$ for BP2 ranges from -12 to -2 fb at the considered parameter space, and its maximum value of -11.93 fb reaches at $\sqrt{s}_{\gamma\gamma}=650\gev$ and $m_{H^0}=250\gev$. For all center-of-mass energies, there appears the dip near $m_{H^0}=420\gev$, which corresponds to $m_{H^\pm}=420\gev$ and $m_{A^0}=340.26\gev$, due to the threshold
effect $m_{H^\pm}\sim m_{W^\pm}+m_{A^0}$. The corresponding relative correction mostly ranges from $-10\%$ to $-35\%$ as seen from the contour lines in Fig.~\ref{fig:BP2}(c). Its magnitude increases with increasing of $\sqrt{s}_{\gamma\gamma}$ and reaches around $-35\%$ at $\sqrt{s}_{\gamma\gamma}=3\tev$ in the region of $250\gev\leq m_{H^\pm}\leq320\gev$. Note that the relative correction becomes positive in a small region due to the production threshold ($\sqrt{s}_{\gamma\gamma}\thickapprox 2 m_{H^\pm}$). For example, at $\sqrt{s}_{\gamma\gamma}=1\tev$, the relative correction increases from $-16.59\%$ to $-7.13\%$ when $m_{H^\pm}$ running from $250\gev$ to $490\gev$. The NLO corrected cross section $\sigma_{\text{LO+NLO}} (\gamma\gamma \rightarrow H^{-}H^{+})$ reaches a maximum value of $130.46$ fb with $\delta_r =-7.23\%$ at $\sqrt{s}_{\gamma\gamma}=550\gev$ for $m_{H^\pm}=250\gev$. At $\sqrt{s}_{\gamma\gamma}=1.5\tev$,  when $m_{H^\pm}$ running from $250$ to $500\gev$, the $\sigma_{\text{LO+NLO}} (\gamma\gamma \rightarrow H^{-}H^{+})$ decreases from $29.99$ to $19.16$ fb while $\delta_r$ changes from $-22.63\%$ to $-18.27\%$.

\begin{figure}[!hbt]
    \begin{center}
\includegraphics[scale=0.39]{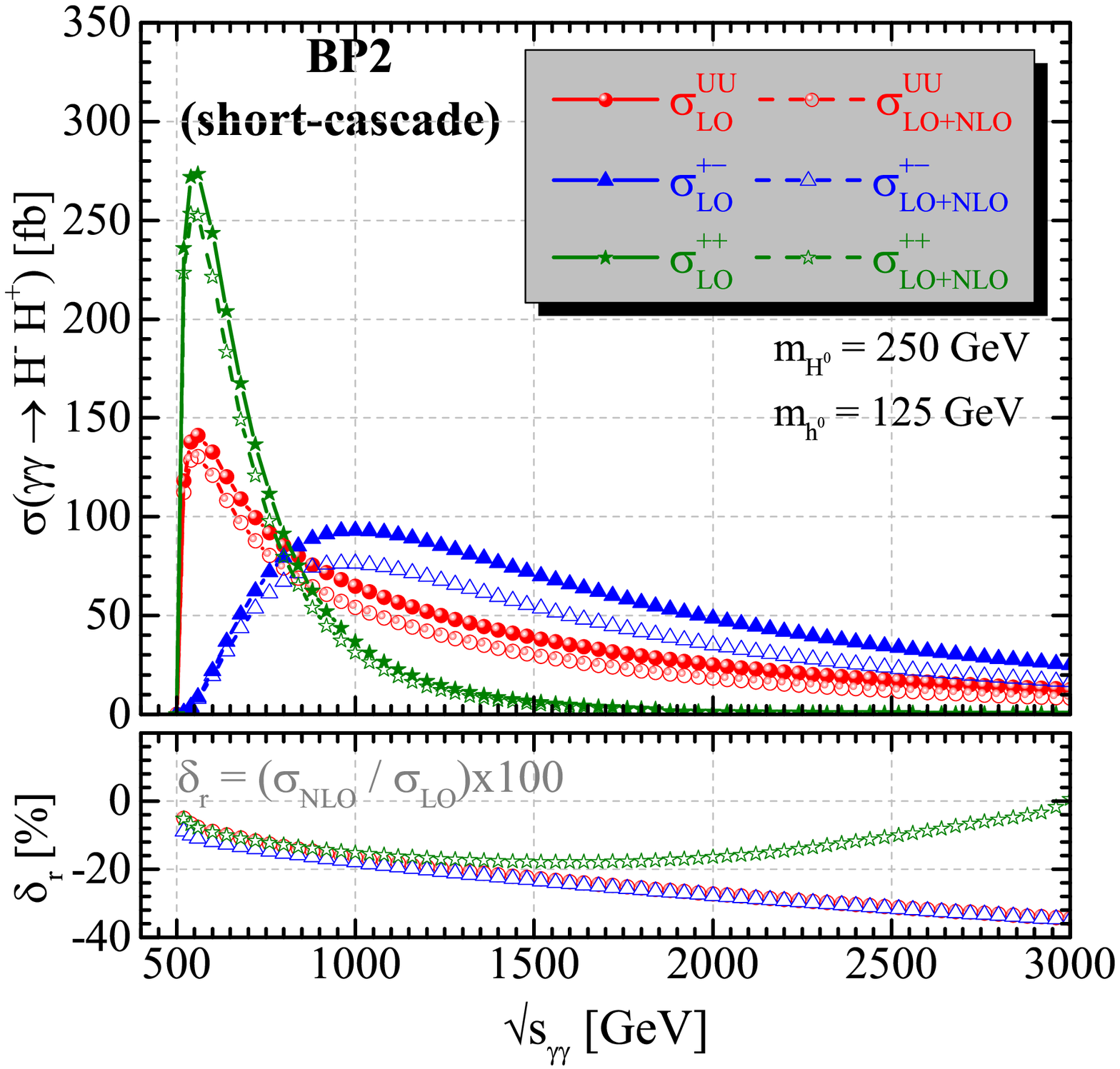}
     \end{center}
     \vspace{ -6mm}
\caption{(color online). The polarized tree-level and full one-loop EW-corrected cross sections of process
$\gamma\gamma \to H^- H^+  $ for different polarization modes of initial beams as a function of $\sqrt{s}_{\gamma \gamma}$ with $m_{H^0}=250\gev$ for BP2.}
\label{fig:BP2pol}
\end{figure}
In Fig.~\ref{fig:BP2pol}, the initial beam polarisation dependence of the integrated tree-level and full one-loop EW-corrected cross sections are plotted as a function of $\sqrt{s}_{\gamma\gamma}$ for BP2, where we take $m_{H^0}=250\gev$. The curves here are like those in Fig.~\ref{fig:BP1pol}. All curves increase firstly, reach their maximal values, and then decrease with the increment of $\sqrt{s}_{\gamma\gamma}$. The $\sigma^{\text{UU}}_{\text{LO}}$ and $\sigma^{\text{UU}}_{\text{LO+NLO}}$ have a peak around $\sqrt{s}_{\gamma\gamma}=560\gev$ with a value of 141.12 fb and 130.33 fb, respectively, with $\delta^{\text{UU}}_r=-7.65\%$. On the other hand, the $\sigma^{+-}_{\text{LO}}$ and $\sigma^{+-}_{\text{LO+NLO}}$ have a peak around $\sqrt{s}_{\gamma\gamma}=1\tev$ with a value of 93.01 fb and 76.29 fb, respectively, with $\delta^{+-}_r=-17.97\%$. The $\sigma^{++}_{\text{LO}}$ and $\sigma^{++}_{\text{LO+NLO}}$ have a peak around $\sqrt{s}_{\gamma\gamma}=540\gev$ with a value of 271.81 fb and 253.17 fb, respectively, with $\delta^{++}_r=-6.86\%$. The relative corrections $\delta^{\text{UU}}_r$ and $\delta^{++}_r$ decrease from $-5\%$ to $-35\%$ while $\delta^{++}_r$ become largest where the production cross section $\sigma^{++}$ goes to zero. When $\sqrt{s}_{\gamma\gamma}$ running from 520 to 3000 GeV, the polarization improvement varies from 0.007 to 1.98 and 2.00 to 0.02 with oppositely polarized photons $(+-)$ and right-handed polarized photons $(++)$, respectively, compared to the unpolarized case.

For BP3, the $\sigma_{\text{NLO}} (\gamma\gamma \rightarrow H^{-}H^{+})$, as seen from Fig.~\ref{fig:BP3}, ranges from -34 to -3 fb at the considered parameter space, and its maximum value of -34.34 fb reaches at $\sqrt{s}_{\gamma\gamma}=150\gev$ and $m_{H^0}=250\gev$. The corresponding relative correction mostly ranges from $-5\%$ to $-30\%$. Its magnitude increases with increasing of $\sqrt{s}_{\gamma\gamma}$ and reaches about $-30\%$ at $\sqrt{s}_{\gamma\gamma}=3\tev$ in the region of $173\gev\leq m_{H^\pm}\leq 317\gev$. Note that the relative correction becomes positive in a small region due to the production threshold ($\sqrt{s}_{\gamma\gamma}\thickapprox 2 m_{H^\pm}$). For example, at $\sqrt{s}_{\gamma\gamma}=1\tev$, the relative correction decreases from $-8.8\%$ to $-14.11\%$ when $m_{H^0}$ running from $250\gev$ to $400\gev$. At $\sqrt{s}_{\gamma\gamma}=0.5\tev$, the $\sigma_{\text{LO}}$ and $\sigma_{\text{LO+NLO}}$ decrease from 436.8 to 111.9 fb and 419.4 to 107.9 fb, respectively, as $m_{H^\pm}$ running from $48.75\gev$ to $243\gev$. The NLO corrected cross section $\sigma_{\text{LO+NLO}} (\gamma\gamma \rightarrow H^{-}H^{+})$ reaches a maximum value of $2.34$ pb with $\delta_r =-1.45\%$ at $\sqrt{s}_{\gamma\gamma}=150\gev$ for $m_{H^\pm}=48.75\gev$. At $\sqrt{s}_{\gamma\gamma}=1\tev$, when $m_{H^0}$ running from $250$ to $500\gev$, the $\sigma_{\text{LO+NLO}} (\gamma\gamma \rightarrow H^{-}H^{+})$ decreases from $111.89$ to $40.28$ fb while $\delta_r$ varies from $-8.8\%$ to $-12.13\%$.
\begin{figure}[!hbt]
    \begin{center}
\includegraphics[scale=0.39]{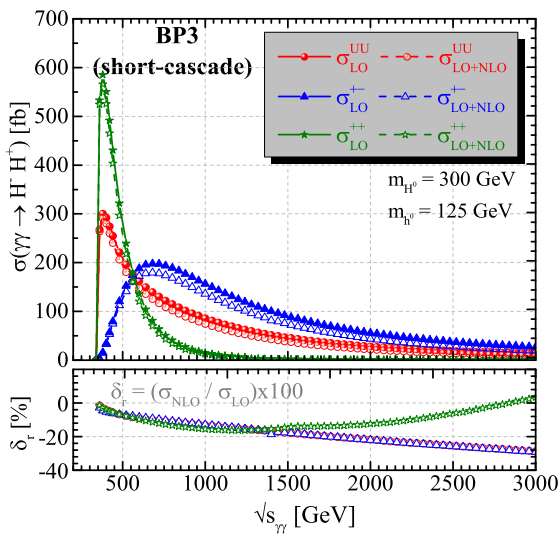}
     \end{center}
     \vspace{-6mm}
\caption{(color online). The polarized tree-level and full one-loop EW-corrected cross sections of process
$\gamma\gamma \to H^- H^+  $ for different polarization modes
of initial beams as a function of $\sqrt{s}_{\gamma \gamma}$ with $m_{H^0}=300\gev$ for BP3.}
\label{fig:BP3pol}
\end{figure}

In Fig.~\ref{fig:BP3pol}, the initial beam polarisation distributions on $\sigma_{\text{LO}}$ and $\sigma_{\text{LO+NLO}}$ are given as a function of $\sqrt{s}_{\gamma\gamma}$ for BP3, where we take $m_{H^0}=300\gev$. The curves here are labeled the same as in Fig.~\ref{fig:BP1pol}. All curves increase firstly, reach their maximal values, and then decrease with the increment of $\sqrt{s}_{\gamma\gamma}$. The $\sigma^{\text{UU}}_{\text{LO}}$ and $\sigma^{\text{UU}}_{\text{LO+NLO}}$ peak around $\sqrt{s}_{\gamma\gamma}=380\gev$ with a value of 300.14 fb and 290.77 fb, respectively, with $\delta^{\text{UU}}_r=-3.12\%$. Moreover, the $\sigma^{+-}_{\text{LO}}$ and $\sigma^{+-}_{\text{LO+NLO}}$ peak around $\sqrt{s}_{\gamma\gamma}=680\gev$ with a value of 197.83 fb and 179.28 fb, respectively, with $\delta^{+-}_r=-9.37\%$. The $\sigma^{++}_{\text{LO}}$ and $\sigma^{++}_{\text{LO+NLO}}$ peak at $\sqrt{s}_{\gamma\gamma}=380\gev$ with a value of 585.02 fb and 565.51 fb, respectively, with $\delta^{++}_r=-3.09\%$.

For BP3, both $\delta^{\text{UU}}_r$ and $\delta^{++}_r$ decrease from $-2\%$ to $-29\%$ while $\delta^{++}_r$ become largest where the production cross section $\sigma^{++}$ goes to zero. At high energies, the integrated cross sections with oppositely polarized photons are enhanced by a factor of 1.99 as compared to the unpolarized case. In the case of both photons with right-handed polarized, the integrated cross sections are highly suppressed at high energies; but at low energies they are amplified up to 1.99 times. The polarization considerably improves the production rate, as expected. This improvement is almost independent of scenarios considered in this study. At other scenarios, similar improvement appears.

\subsection{Low-$m_H$ scenario}
\begin{figure*}[!hbt]
    \begin{center}
\includegraphics[scale=0.39]{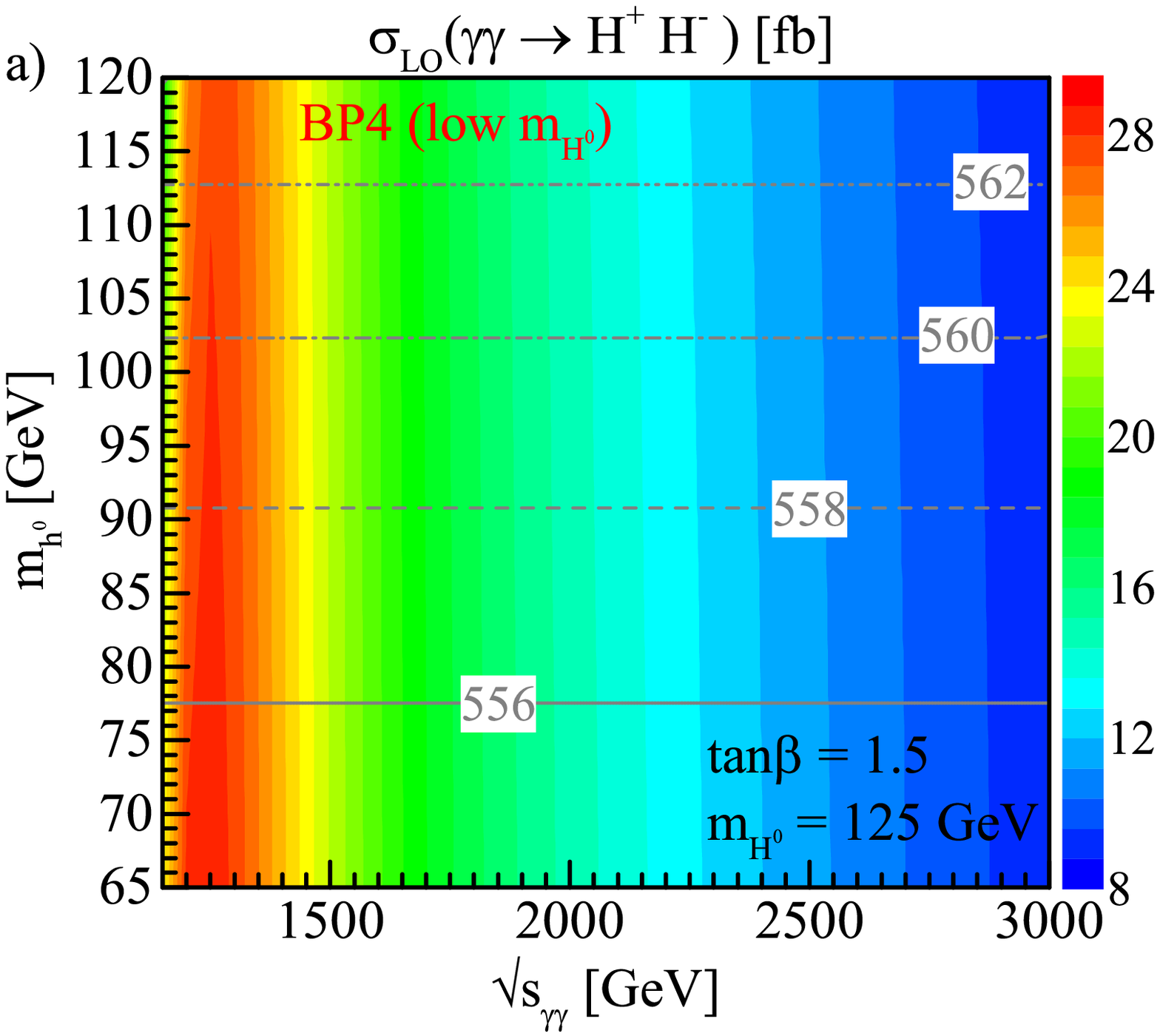}
\includegraphics[scale=0.39]{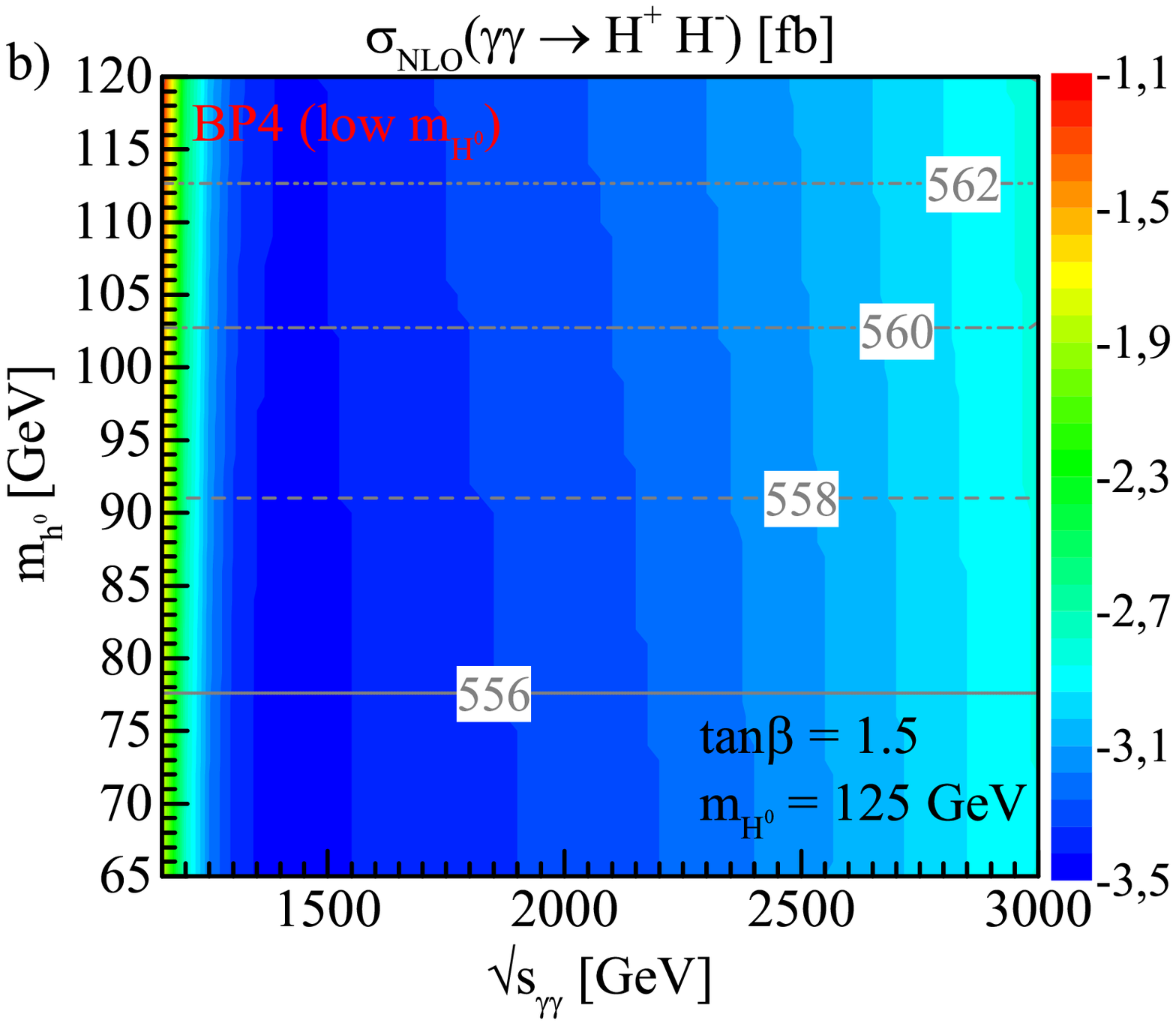}
\includegraphics[scale=0.39]{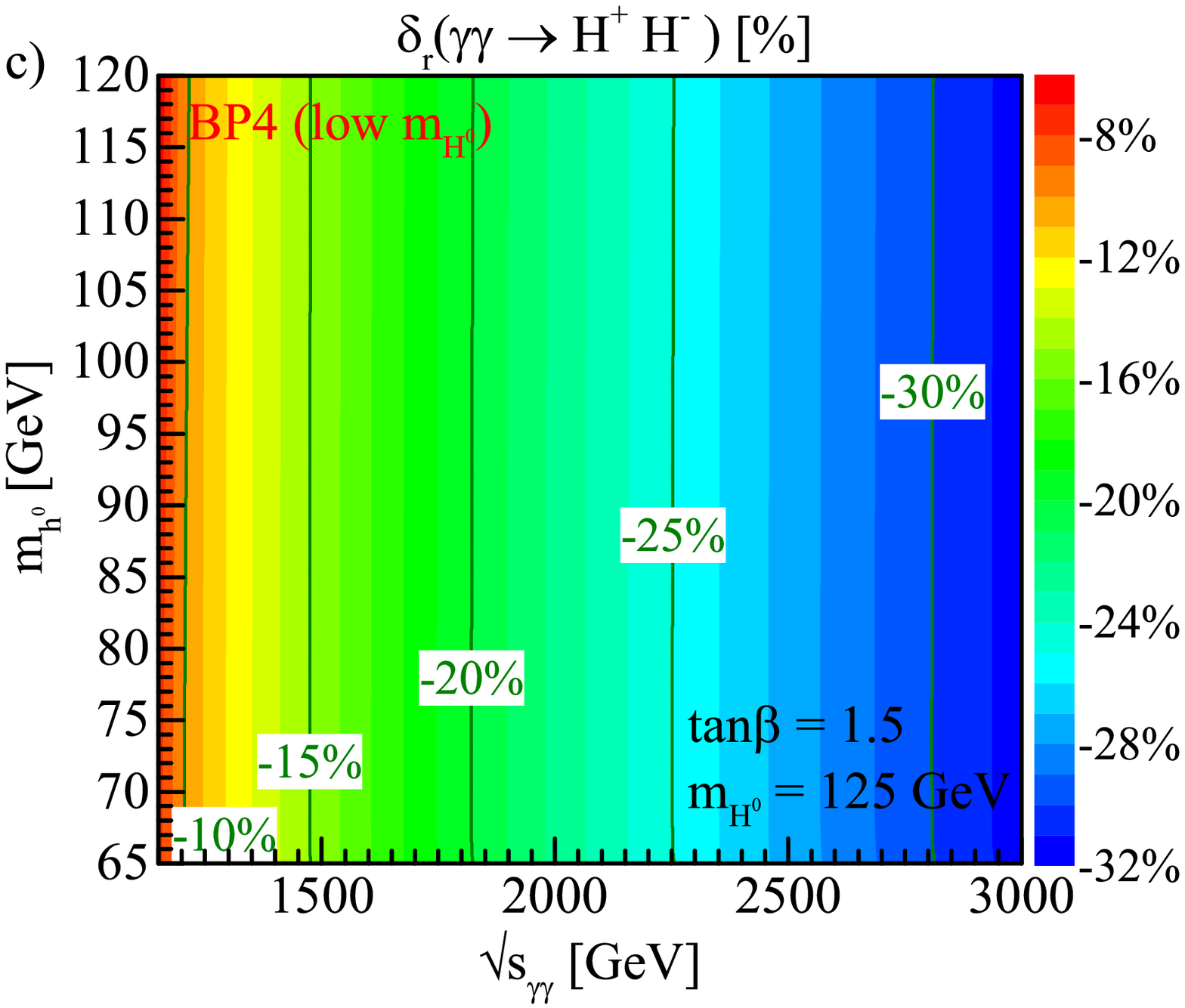}
     \end{center}
     \vspace{ -6mm}
\caption{(color online).  (a) The tree-level and (b) NLO corrections of cross sections (in fb), and (c) the corresponding relative correction for process $ \gamma \gamma \to H^- H^+$ scanned over the ($m_{h^0}$, $\sqrt{s}_{\gamma\gamma}$) plane in the low-$m_H$ scenario. The contour lines correspond to predictions for $m_{H^\pm}$ in unit of GeV, and the relative corrections as a percentage.}
\label{fig:BP4}
\end{figure*}

In Fig.~\ref{fig:BP4}, the tree-level and the NLO corrections of cross sections, and the corresponding relative correction of process $\gamma\gamma \to H^- H^+ $ are scanned over the regions of $m_{h^0}$-$\sqrt{s}$ in the low-$m_H$ scenario, where the heaviest CP-even Higgs $H^0$ behaves like the SM Higgs boson and its mass is taken as $m_{H^0}=125\gev$. The scan parameters are varied as follows: $65 \leq m_{h^0} \leq 120\gev$ in steps of 1 GeV, and $1150 \leq \sqrt{s} \leq 3000\gev$ in steps of 50 GeV.
\begin{figure}[!hbt]
    \begin{center}
\includegraphics[scale=0.39]{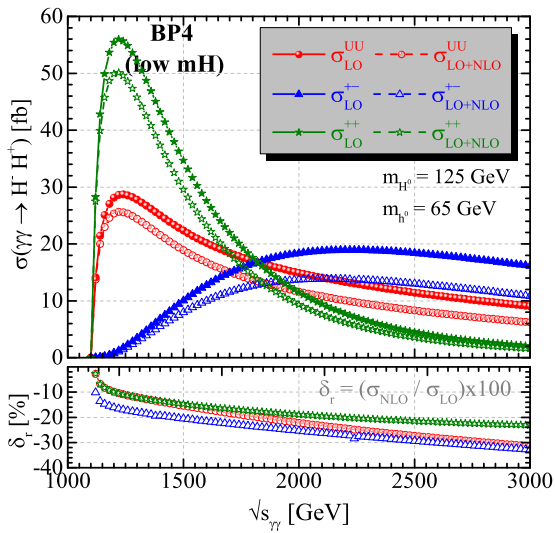}
     \end{center}
     \vspace{ -6mm}
\caption{(color online). The polarized tree-level and full one-loop EW-corrected cross sections of process
$\gamma\gamma \to H^- H^+  $ for different polarization modes
of initial beams as a function of $\sqrt{s}_{\gamma \gamma}$ with $m_{h^0}=65\gev$ for BP4 in the low-$m_H$ scenario.}
\label{fig:BP4pol}
\end{figure}
Similar to the non-alignment scenario, the two Higgs bosons, $A^0$ and $H^\pm$, decouple sufficiently such that they do not affect the phenomenology. The corresponding mass hierarchy is $m_{h^0}<m_{H^0}=125\gev<m_{H^\pm}=m_{A^0}$. The values of the mass of charged Higgs boson $m_{H^\pm}$, which are calculated in terms of the mass of neutral Higgs boson $m_{h^0}$, are shown with contour lines, and the $m_{H^\pm}$ changes very slowly with increasing of the $m_{h^0}$. When $m_{h^0}$ runs from 65 to 120 GeV, $m_{H^\pm}$ varies from $554$ to $563\gev$ for BP4. Due to being $c_{\beta-\alpha}=1$ in the low-$m_H$ scenario, the coupling $C_{H^0 H^- H^+}$ would reach its largest value affecting the cross section. However, the effect of $C_{h^0 H^- H^+}$ on the total cross-section will be reduced.

The integrated cross sections $\sigma_{\text{LO}} (\gamma\gamma \rightarrow H^{-}H^{+})$ and $\sigma_{\text{NLO}} (\gamma\gamma \rightarrow H^{-}H^{+})$ change very slowly with $m_{h^0}$, due to the small range of mass $m_{H^\pm}$. They decrease with increasing of $\sqrt{s}_{\gamma\gamma}$ when $m_{H^\pm} \ll \sqrt{s}/2$. Particularly, they reaches its larger values for $\sqrt{s}_{\gamma\gamma}< 1500\gev$ in the scan region. The NLO corrections make negative contributions to total cross section as in the previously discussed scenarios. The $\sigma_{\text{NLO}} (\gamma\gamma \rightarrow H^{-}H^{+})$ ranges from -3.5 to -3 fb at most of the parameter space. On the other hand, the relative correction mostly ranges from $-10\%$ to $-32\%$ as seen from the contour lines in Fig.~\ref{fig:BP4}(c). Its magnitude increases with increasing of $\sqrt{s}_{\gamma\gamma}$. For example, at $\sqrt{s}_{\gamma\gamma}=1.2\tev$, the relative correction increases from $-9.85\%$ to $-9.53\%$ when $m_{h^0}$ running from $65\gev$ to $120\gev$. The NLO corrected cross section $\sigma_{\text{LO+NLO}} (\gamma\gamma \rightarrow H^{-}H^{+})$ reaches a maximum value of $25.47$ fb with $\delta_r =-11.1\%$ at $\sqrt{s}_{\gamma\gamma}=1250\gev$ for $m_{H^\pm}=554\gev$. Overall, the $\sigma_{\text{LO+NLO}} (\gamma\gamma \rightarrow H^{-}H^{+})$ appears usually in the range of 6 to 25 fb for considered parameter regions of the low-$m_H$ scenario.

Figure~\ref{fig:BP4pol} presents the initial beam polarisation dependence of the integrated tree-level and full one-loop EW-corrected cross sections versus $\sqrt{s}_{\gamma\gamma}$ for BP4, where we take $m_{h^0}=65\gev$. The $\sqrt{s}_{\gamma\gamma}$ varies from the value little larger than the threshold $2 m_{H^\pm}$ to $3\tev$. The curves here are labeled the same as in Fig.~\ref{fig:BP1pol}. It is seen that all curves increase firstly, reach their maximal values, and then decrease with the increment of $\sqrt{s}_{\gamma\gamma}$. The $\sigma^{\text{UU}}_{\text{LO}}$ and $\sigma^{\text{UU}}_{\text{LO+NLO}}$ have a peak around $\sqrt{s}_{\gamma\gamma}=1220\gev$ with a value of 28.6 fb and 25.63 fb, respectively, with $\delta^{\text{UU}}_r=-10.4\%$. Moreover, the $\sigma^{+-}_{\text{LO}}$ and $\sigma^{+-}_{\text{LO+NLO}}$ have a peak around $\sqrt{s}_{\gamma\gamma}=2200\gev$ with a value of 18.91 fb and 13.79 fb, respectively, with $\delta^{+-}_r=-27.07\%$. The $\sigma^{++}_{\text{LO}}$ and $\sigma^{++}_{\text{LO+NLO}}$ have a peak at $\sqrt{s}_{\gamma\gamma}=1220\gev$ with a value of 55.96 fb and 50.09 fb, respectively, with $\delta^{++}_r=-10.49\%$.

On the other hand, the absolute relative corrections increase with the increment of $\sqrt{s}_{\gamma\gamma}$ for all polarization cases. The relative corrections $\delta_r$ change in the ranges of $\delta^{\text{UU}}_r \in [-2.8\%, -31.4\%]$,  $\delta^{+-}_r \in [-10.1\%, -32.9\%]$ and  $\delta^{++}_r \in [-2.8\%, -23.2\%]$, when the $\sqrt{s}_{\gamma\gamma}$ goes from $1120\gev$ to $3\tev$. At high energies, the  $\sigma^{+-}_{\text{LO}}$ and  $\sigma^{+-}_{\text{NLO}}$ are enhanced by a factor of 1.7 as compared to the unpolarized case. The  $\sigma^{++}_{\text{LO}}$ and  $\sigma^{++}_{\text{NLO}}$ are significantly suppressed for the region of $ \sqrt{s}_{\gamma\gamma}> 2.5\tev$ ; but for the region of $ 1.1 < \sqrt{s}_{\gamma\gamma}< 1.5\tev$ they are amplified by between 1.5 and 2 times. It can be seen that the longitudinal polarization of initial photons increases the production rate of $H^- H^+$ signal in the photon-photon colliders.

\subsection{Angular Distribution of the Differential Cross Section}
\begin{figure}[!hbt]
    \begin{center}
\includegraphics[scale=0.41]{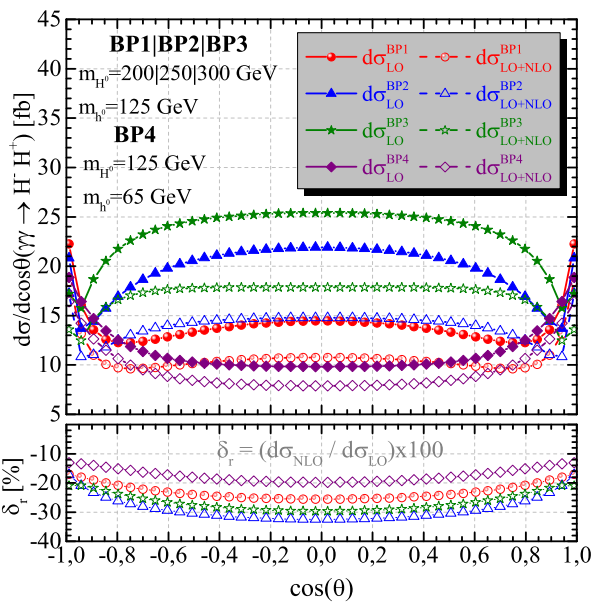}
     \end{center}
     \vspace{ -6mm}
\caption{(color online). The differential cross section versus $\cos{\theta}$ at $\sqrt{s}_{\gamma\gamma}=1.5\tev$.}
\label{fig:BPsdifcs}
\end{figure}
The tree-level and the virtual plus soft photon corrected of differential cross sections of $ \gamma \gamma \rightarrow H^{-}H^{+}$ are presented in Fig.~\ref{fig:BPsdifcs} as a function of the angle between the initial photon and the charged Higgs boson at $\sqrt{s}_{\gamma\gamma}=1.5\tev$. Also, in the same figure, the corresponding relative corrections as a function of $\cos{\theta}$ are shown on the bottom panel for each BPs. It can be seen in this figure that all curves are rather symmetric according to $\cos({\theta})=0$. The tree-level differential cross sections relatively flat, particularly in the region of $-0.6 <\cos{\theta}<0.6$, but the virtual plus soft photon corrections significantly depend on the angle $\theta$. The corrections reach their maximum values when $\cos{\theta}$ have values of +1 and -1. Namely, the charged Higgs pairs are dominantly produced in the backward and forward directions and it will be much more possible to detect them
in that region of the collision. On the other hand, the differential cross-section is smaller in large values of the charged Higgs mass. When $\cos{\theta}$ goes from 0 to +1 or -1, the relative correction $\delta_r$  varies from $-25.58\%$ to $-17.01\%$ for BP1, $-32.48\%$ to $-17.21\%$ for BP2, $-29.67\%$ to $-20.96\%$ for BP3, and $-19.80\%$ to $-13.08\%$ for BP4.

\subsection{Process  $e^- e^+ \rightarrow \gamma \gamma \rightarrow H^{-}H^{+}$}
\begin{figure}[!b]
    \begin{center}
\includegraphics[scale=0.43]{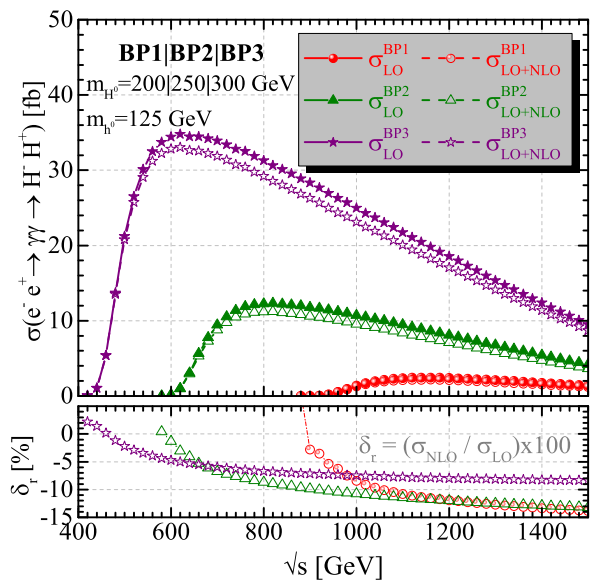}
     \end{center}
     \vspace{ -6mm}
\caption{(color online). The tree-level and full one-loop level EW-corrected cross-sections (in fb) convoluted with the photon luminosity of the parent process $e^+e^-\rightarrow\gamma\gamma\rightarrow H^+ H^-$ versus $\sqrt{s}$ for BP1, BP2 and BP3.}
\label{fig:BPsee}
\end{figure}
The tree-level and the full one-loop level EW-corrected cross sections of the parent process $e^- e^+ \rightarrow \gamma \gamma \rightarrow H^{-}H^{+}$ obtained by convoluting with the luminosity of photon are presented in Fig.~\ref{fig:BPsee} as a function of $e^- e^+$ center-of-mass energy. Also, in the same figure, the corresponding relative correction as a function of $\sqrt{s}$ is shown on the bottom panel. From this figure we find the expected behavior: a rapid increase near the production threshold, followed by a decrease with the increment of the colliding $e^- e^+$ center-of-mass energy. It is obvious that the production rate in BP3 is always larger than those in the other BPs. The virtual plus the real corrections, i.e., NLO corrections are mostly negative for each BPs. For non-alignment scenario, the $\sigma^{\text{BP1}}_{\text{LO}}$ and $\sigma^{\text{BP1}}_{\text{LO+NLO}}$ reach their maximum values of 2.45 fb and 2.17 fb at $\sqrt{s}=1160\gev$, respectively, with $\delta^{\text{BP1}}_r=-11.56\%$. For BP2 in short-cascade scenario, the $\sigma^{\text{BP2}}_{\text{LO}}$ and $\sigma^{\text{BP2}}_{\text{LO+NLO}}$ reach their maximum values of 12.34 fb and 11.24 fb at $\sqrt{s}=820\gev$, respectively, with $\delta^{\text{BP2}}_r=-8.94\%$. For BP3 in short-cascade scenario, the $\sigma^{\text{BP3}}_{\text{LO}}$ and $\sigma^{\text{BP3}}_{\text{LO+NLO}}$ reach their maximum values of 34.78 fb and 33.03 fb at $\sqrt{s}=620\gev$, respectively, with $\delta^{\text{BP3}}_r=-5.03\%$.
For all BPs, it is clear that the absolute relative correction increases with the increasing of $\sqrt{s}$.
For BP1, the $\delta_r$  varies from $-2.78\%$ to $-13.88\%$ as $\sqrt{s}$ goes from 900 GeV to 1.5 TeV. For BP2, the $\delta_r$ varies from $-1.32\%$ to $-13.21\%$ as $\sqrt{s}$ goes from 600 GeV to 1.5 TeV. Finally, for BP3, the $\delta_r$ ranges from $+1.40\%$ to $-8.37\%$ when $\sqrt{s}$ runs from 440 GeV to 1.5 TeV.

\subsection{Decay channels of the charged Higgs boson}
The final decay products of the produced charged Higgs bosons will be analyzed for all scenarios in this section. The decay channels are calculated by using {\tt 2HDMC 1.7.0}. To explore the process in a collider, we must firstly identify all the possible charged Higgs products.
\begin{figure}[!b]
    \begin{center}
\includegraphics[scale=0.43]{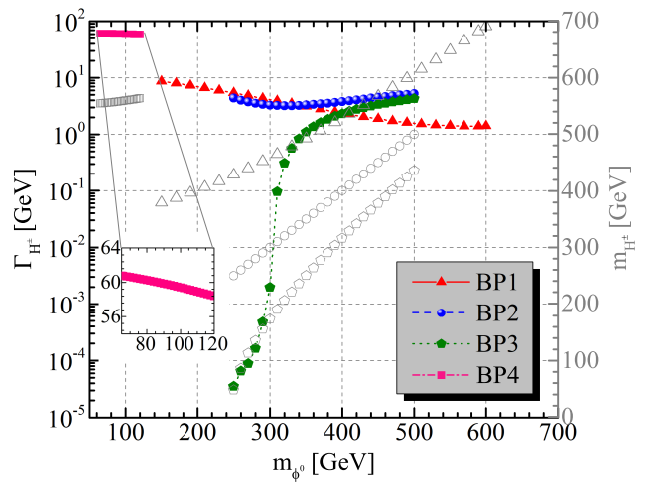}
     \end{center}
     \vspace{-5mm}
\caption{(color online). The total decay widths of charged Higgs boson $H^\pm$ predicted by the scenarios discussed in the text.}
\label{fig:TotalWidht}
\end{figure}
\begin{figure*}[!hbt]
    \begin{center}
\includegraphics[scale=0.4]{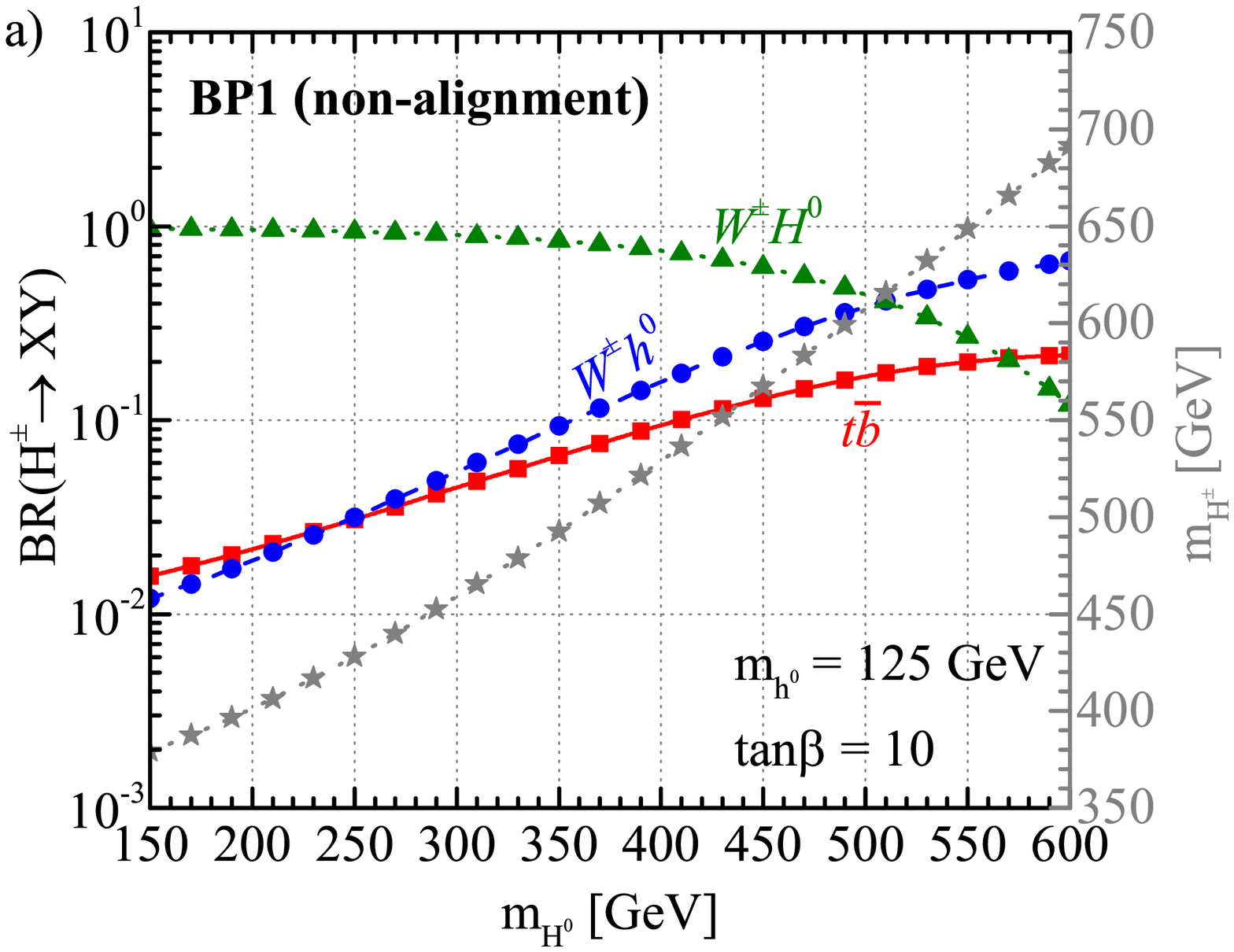}
\includegraphics[scale=0.4]{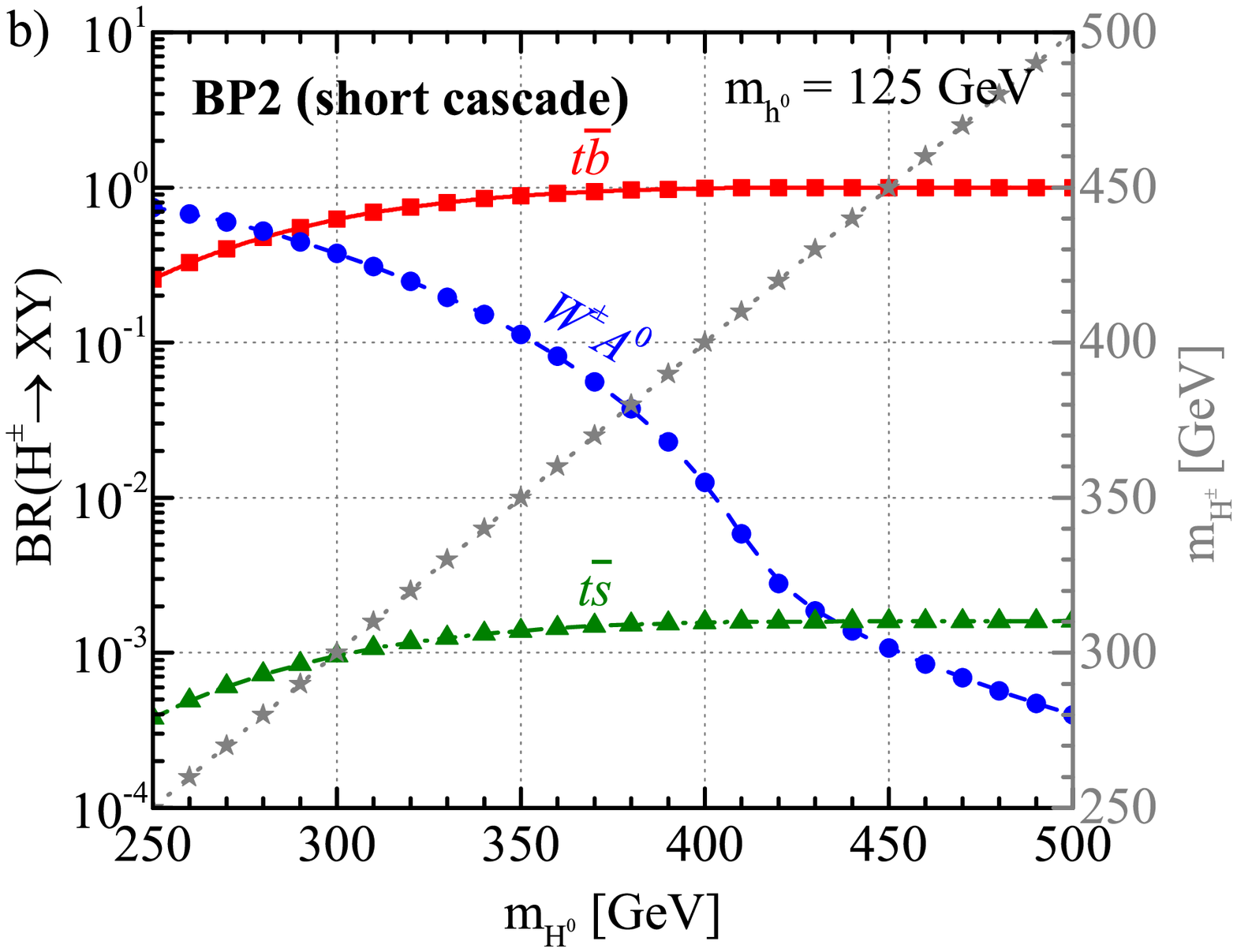}
\includegraphics[scale=0.4]{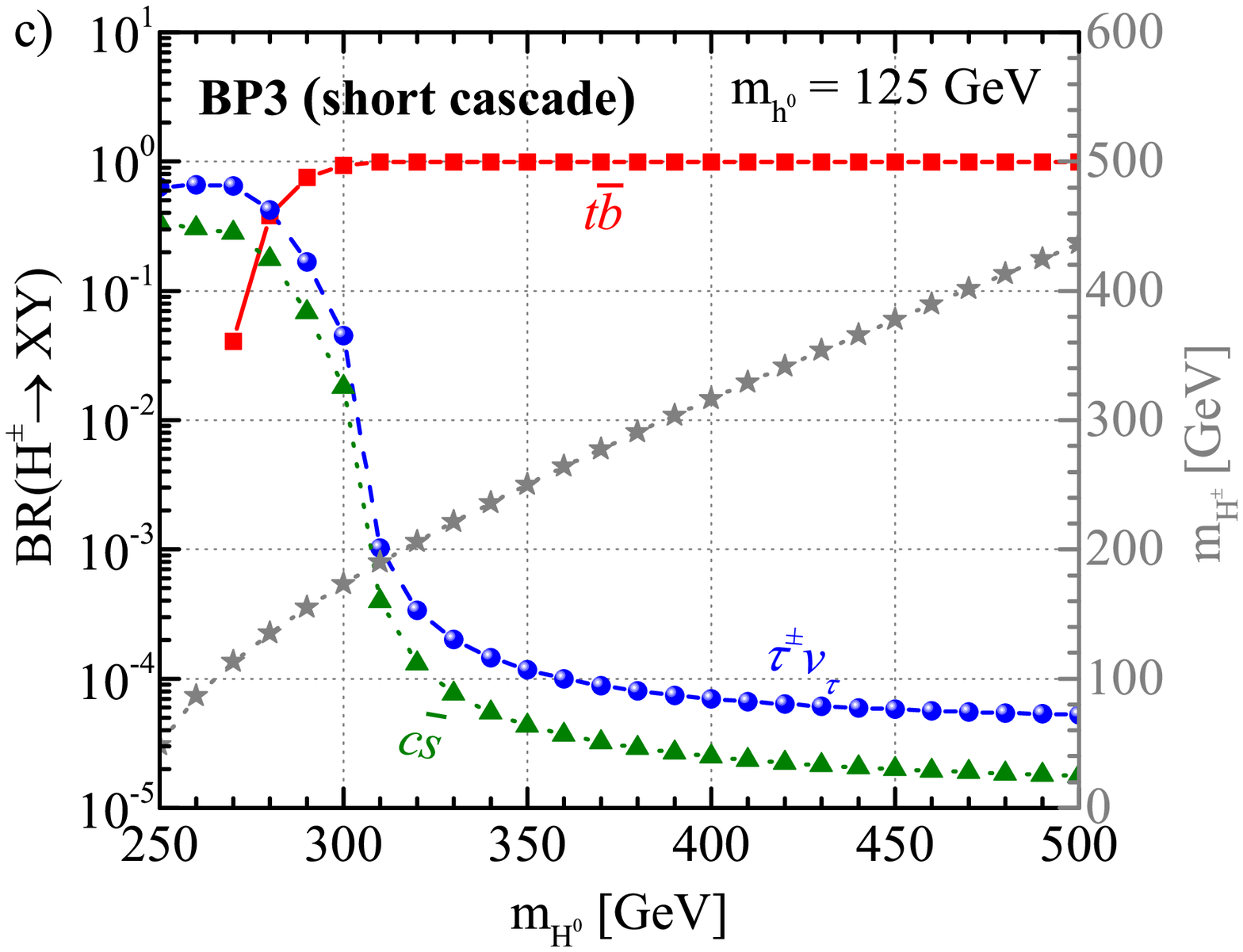}
\includegraphics[scale=0.4]{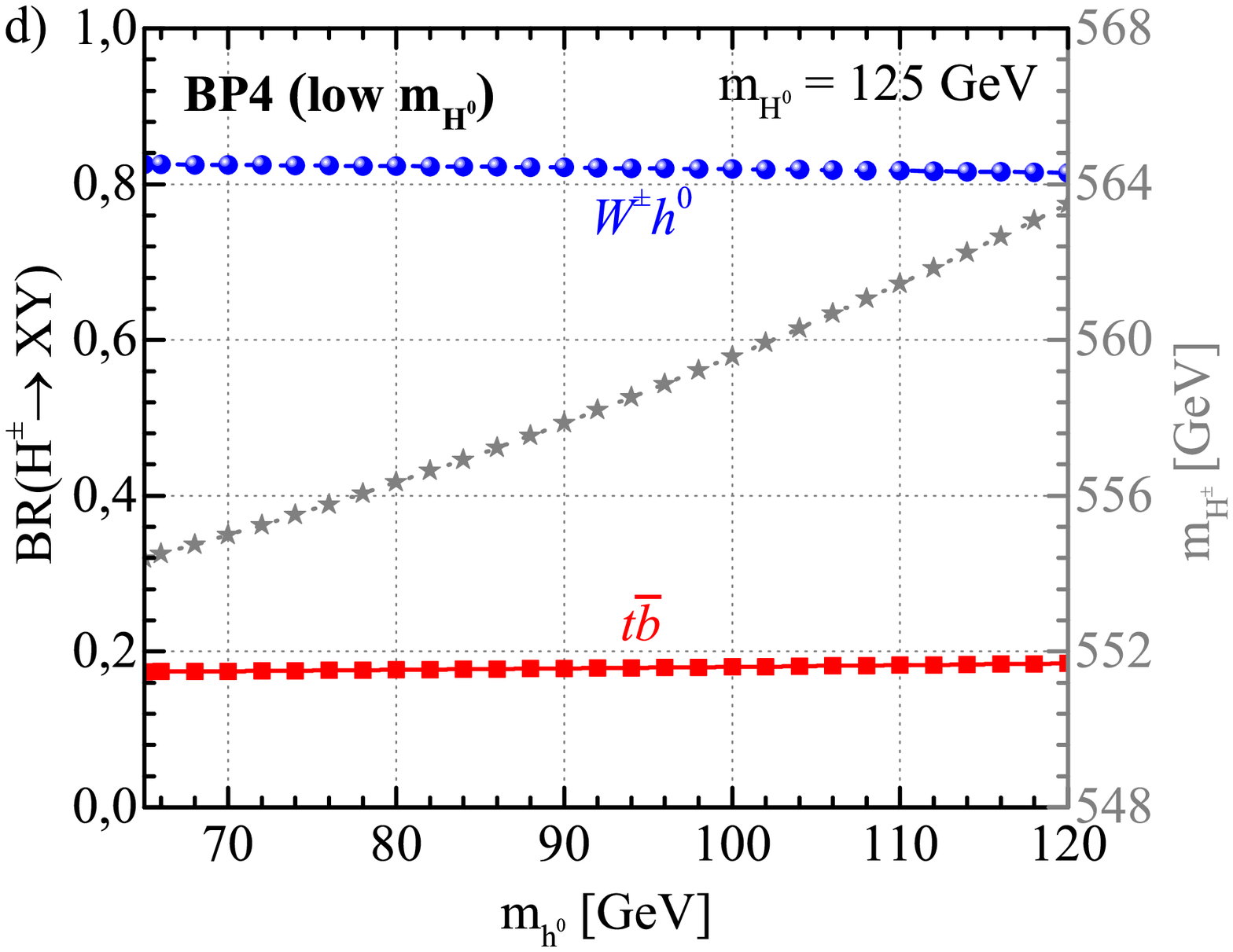}
     \end{center}
     \vspace{ -5mm}
\caption{(color online). The branching ratios of charged Higgs boson $H^\pm$  predicted by the scenarios discussed in the text. Other modes for which BR value is less than $10^{-4}$ are omitted here for clarity.}
\label{fig:BRs}
\end{figure*}
The total decay widths of charged Higgs boson $H^\pm$ versus the mass of Higgs boson, $m_{\phi^0}$ where $\phi^0$ is $h^0$ for low-$m_{H^0}$ scenario and $H^0$ for other two scenarios, are plotted in Fig.~\ref{fig:TotalWidht} for all scenarios considered in this study. Also, in the same figure, the mass of charged Higgs boson as a function of $m_{\phi^0}$ is shown by hollow symbols on the right axis. In all scenarios, the charged Higgs boson mass increases with increment of the neutral Higgs mass $m_{\phi^0}$, as expected. This figure shows that the decay widths are quite sensitive to the charged Higgs mass and the mass hierarchy. The decay width decreases when the $m_{H^0}-m_{H^\pm}$ mass splitting is small. For BP1, when $m_{H^\pm}$ runs from 379 to 691 GeV, $\Gamma_{H^\pm}$ decreases from 8.66 to 1.42. However, for BP2, $\Gamma_{H^\pm}$ increases from 4.38 to 5.22 in the mass window $250\gev < m_{H^\pm} < 500\gev$. For BP3, $\Gamma_{H^\pm}$ increases from 3.5$\times10^{-5}$ to 4.28 as $m_{H^\pm}$ increases from $48.75\gev$ to $436\gev$. For BP4 in the low-$m_H$ scenario, $\Gamma_{H^\pm}$ decreases from 60 to 58 when $m_{H^\pm}$ changes from $558$ to $564\gev$.

In Fig.~\ref{fig:BRs}, we show the branching ratios of charged Higgs boson for the dominant decay modes as a function of $m_{h^0, H^0}$ for all scenarios discussed in the text. The mass of charged Higgs boson as a function of $m_{h^0, H^0}$ is also shown by star symbols on the right axis. Note that the decay modes for which BR value is less than $10^{-4}$ are omitted here for clarity. When considering only decays to SM particles, the dominant decay modes are those involving the heaviest lepton or quark pairs accessible such as follows: $\tau \nu_{\tau}$, $t\bar{b}$, $t\bar{s}$ and $c\bar{s}$. Among these, the decay mode $t\bar{b}$ of the charged Higgs boson is available in all BPs significantly with varying branching ratios. In scenarios with Type I, due to the $\cot\beta$ dependence of coupling, the $Br(H^\pm \to \tau^\pm \nu_{\tau})$ is suppressed by $m_\tau^2/m_t^2$ over $Br(H^\pm \to t\bar{b})$.

In the non-alignment scenario, the dominant decay mode of charged Higgs is in the $W^\pm H^0$ channel, following sub-dominant channels  $t\bar{b}$ and $W^\pm h^0$ for BP1, followed by other suppressed modes such as, $H^\pm \to t\bar{s}$ and $c\bar{s}$, particularly in the range $m_{H^0}<500\gev$. Then, $Br(H^\pm \to W^\pm H^0)$ gradually decreases at larger values of $m_{H^0}$, and the decay mode $W^\pm h^0$ becomes dominant. $Br(H^\pm \to W^\pm h^0)$ increases from 1.2 to $66.3\%$ in the mass window $150\gev < m_{H^0} < 600\gev$. For $m_{H^0}<500\gev$, the process becomes $\gamma\gamma \to H^+ H^- \to W^+ H^0 W^- H^0$. On the other hand, the dominant decay mode of $H^0$ is $W^+W^-$ with branching ratio of $88.7-50.2\%$. When the hadronic decays of $W^\pm$-boson are considered, there appear 12 jets at the final state. Consequently, it is difficult to reconstruct the $ W^\pm$, so the charged Higgs bosons.

In the short-cascade scenario, once the decay $H^\pm \to t\bar{b}$ opens up (when $m_{H^\pm}> m_t + m_b$), it quickly becomes dominant, leading almost $100\%$ Br. On the other hand, the decay modes $W^\pm A^0$ and $t\bar{s}$ ($\tau^\pm \nu_\tau$ and $c\bar{s}$ ) get suppressed for $m_{H^0}>300\gev$ in the BP2 (BP3). In the short-cascade scenario, the process becomes $\gamma\gamma \to H^+ H^-  \to tb tb$, and the decays of $t$ may be an ideal option for reconstructing the process at $m_{H^0}>300\gev$. The subsequent decays of $t \to Wb, W\to q\bar{q}(l\nu_l)$ will form the signature of $H^\pm$ at a detector. Consequently, it can be tagged with 8-jets plus 2-$b$-tagged jets for the short-cascade scenario.

In the low-$m_H$ scenario, the decay channel $H^\pm \to W^\pm h^0$ is clearly dominant over the full mass range ($65\gev < m_{h^0} < 120\gev$), because its decay width is proportional to $c_{\beta-\alpha}$ leading it dominant $(\sim 100\%)$ for the choice of $s_{\beta-\alpha}=0$. The sub-dominant channel for BP4 is $H^\pm \to t\bar{b}$ with around $18\%$ Br.
The $Br(h^0 \to  b \bar{b})$ changes between $90.0-87.7\%$. Therefore, the process can be tagged with 4-jets plus 4-$b$-tagged jets.

\section{Summary and Conclusions}\label{sec:conc}
The 2HDM is the simplest extension of the SM which contains the charged Higgs boson. In the case of a discovery of charged Higgs boson, a subsequent exact measurement of its properties will be important for determining its nature and the corresponding model parameters. In order to provide enough precision, full one-loop contributions need to be included in the production channels of charged Higgs boson. The pair production of charged Higgs is one of the main channels that would provide an observable signal in a wide range of the parameter space in 2HDM. In this study, the charged Higgs pair production has been studied via $\gamma \gamma$ collisions, considering a complete set of one-loop EW corrections in the framework of 2HDM. In the one-loop diagrams, the UV divergences have been regularized by dimensional regularization in the on-mass-shell renormalization scheme, and IR divergences have been canceled by the inclusion of soft and hard QED radiation. The numerical evaluation was carried out for three different scenarios, so-called non-alignment, low-$m_H$ and short-cascade, defined in the framework of 2HDM, in the presence of the up-to-date experimental constraints. The tree-level and full one-loop EW corrections of total cross sections have been scanned over the plane ($m_{\phi^0},\sqrt{s}$), where $\phi^0$ is $h^0$ for low-$m_{H^0}$ scenario and $H^0$ for other two scenarios (non-alignment and short-cascade scenarios). The regions of the parameter space in which the production rates including the relative one-loop corrections are sufficiently large have been highlighted for each scenario.

The results show that the one-loop EW corrections mostly reduce the tree-level cross section and the relative correction is typically few tens of percent for both the $\gamma\gamma\rightarrow H^{-}H^{+}$ and the $e^{-}e^{+} \rightarrow  \gamma\gamma\rightarrow H^{-}H^{+}$ depending on chosen parameter space. The virtual plus the real corrections are mostly negative for selected BPs. The overall effect range between $-10\%$ and $-30\%$ in a wide range of the model parameters. Since the tree-level cross-section is mostly from QED, the model-dependent parameters appear firstly at one-loop level. The cross section in the short-cascade scenario is always larger than those in the other two scenarios and the cross section of the low-$m_{H^0}$ scenario is the smallest one among all of the three scenarios. The production rate with light charged Higgs bosons is larger than that with heavy charged Higgs bosons owing to the larger final state phase space volume for either an electron-positron or photon-photon collider. The full one-loop corrected cross sections of $\gamma\gamma \rightarrow H^{-}H^{+}$ reach their maximum values as follows: $\sigma^{\text{BP1}}_{\text{LO+NLO}}=51.97$ fb with $\delta_r =-11.28\%$, $\sigma^{\text{BP2}}_{\text{LO+NLO}}=130.46$ fb with $\delta_r =-7.23\%$, $\sigma^{\text{BP3}}_{\text{LO+NLO}}= 2.34$ pb with $\delta_r =-1.45\%$ and $\sigma^{\text{BP4}}_{\text{LO+NLO}}=25.47$ fb with $\delta_r =-11.1\%$. The absolute relative corrections increase with the increment of $\sqrt{s}$ for all cases.

The production rates of $\gamma\gamma\rightarrow H^{-}H^{+}$ in different polarization collision modes of initial beams have been also discussed. The production rate of $\gamma\gamma \rightarrow H^{-}H^{+}$ is enhanced up to around 2 times with oppositely polarized photons at high energies and right-handed polarized photons at low energies, as independent of the scenarios interested. Consequently, having both photons polarized can turn out to be significant
to ensure a measurable production rate.

The reconstruction of the charged Higgs boson has been presented for each scenarios, studying its dominant decay modes.
In the non-alignment and low-$m_H$ scenarios, the bosonic decay channels $H^\pm \to W^\pm H^0$ and $ W^\pm h^0$ are dominant, respectively, while in the short-cascade scenario, the dominant decay channel is $H^\pm \to t\bar{b}$. The bosonic decay channels are highly suppressed due to alignment limit and limited phase space.

In summary, the first phenomenological results in the context of 2HDM for the one-loop EW corrections to the charged Higgs pair production via photon-photon collisions have been produced, and  in the light of this, the main distinctive features between the selected scenarios have been highlighted. The precise measurements for the associated production would be possible at the future colliders, and our results will be helpful for determining new physics signals based on the 2HDM and putting more precise limits on the model parameters.

\end{document}